\documentclass[aps, a4paper, twocolumn, nofootinbib, showpacs, amsfonts]{revtex4}

\usepackage[utf8]{inputenc}
\usepackage{dsfont}
\usepackage{amssymb}
\usepackage{graphicx}
\usepackage{amsmath}
\usepackage{bm}
\usepackage{xcolor}
\usepackage{mathtools}
\usepackage{tabularx}
\usepackage{lipsum}
\usepackage{amsmath}
\usepackage{booktabs}
\usepackage{hyperref}
\usepackage{multirow}

\newcommand{\bea}{\begin{eqnarray}}
\newcommand{\eea}{\end{eqnarray}}
\newcommand{\be}{\begin{equation}}
\newcommand{\ee}{\end{equation}}
\newcommand{\mplus}{m_{+}}
\newcommand{\mminus}{m_{-}}
\newcommand{\cM}{\mathcal{M}}

\usepackage{tikz}
\usetikzlibrary{decorations.pathmorphing}

\newcommand{\p}{\partial}

\begin{document}

\title{Cauchy Horizon (In)Stability of Regular Black Holes}

\author{Alfio Bonanno}
\email{alfio.bonanno@inaf.it}
\affiliation{INAF, Osservatorio Astrofisico di Catania, via S.Sofia 78, I-95123 Catania, Italy}
\affiliation{INFN, Sezione di Catania, via S.Sofia 64, I-95123 Catania, Italy}

\author{Antonio Panassiti}
\email{antonio.panassiti@phd.unict.it}
\affiliation{INAF, Osservatorio Astrofisico di Catania, via S.Sofia 78, I-95123 Catania, Italy}
\affiliation{INFN, Sezione di Catania, via S.Sofia 64, I-95123 Catania, Italy}
\affiliation{Dipartimento di Fisica e Astronomia “Ettore Majorana”, Universit\`a di Catania, Via S.
Sofia 64, 95123, Catania, Italy}
\affiliation{High Energy Physics Department, Institute of Mathematics, Astrophysics, and Particle Physics, Radboud University, Heyendalseweg 135, 6525 AJ Nijmegen, The Netherlands}

\author{Frank Saueressig}
\email{f.saueressig@science.ru.nl}
\affiliation{High Energy Physics Department, Institute of Mathematics, Astrophysics, and Particle Physics, Radboud University, Heyendalseweg 135, 6525 AJ Nijmegen, The Netherlands}

\begin{abstract}
A common feature of regular black hole spacetimes is the presence of an inner Cauchy horizon. The analogy to the Reissner-Nordstr{\"o}m solution then suggests that these geometries suffer from a mass-inflation effect, rendering the Cauchy horizon unstable. Recently, it was shown that this analogy fails for certain classes of regular black holes, including the Hayward solution, where the late-time behavior of the mass function no longer grows exponentially but follows a power law. In this work, we extend these results in a two-fold way. First, we determine the basin-of-attraction for the power-law attractor, showing that the tamed growth of the mass function is generic. Second, we extend the systematic analysis to the Bardeen geometry, the Dymnikova black hole, and a spacetime arising from a non-singular collapse model newly proposed in the context of asymptotically safe quantum gravity. Remarkably, in the latter solution, the Misner-Sharp mass at the Cauchy horizon remains of the same order of magnitude of the mass of the black hole, since its growth is just logarithmic.
\end{abstract}

\maketitle

\section{Introduction}
\label{Sec.Intro}
Black holes are among the most fascinating objects in theoretical physics \cite{Hawking:1973uf, Buoninfante:2024oxl}. Arguably, a thorough understanding of these objects may the key towards understanding the working of gravity in the strong gravity regime or at the level of a quantum theory. Within classical general relativity black holes are rather simple objects. Essentially, the corresponding stationary solutions consist of a spacetime singularity which is hidden inside an (outer) event horizon. Once charge is added, the Schwarzschild solution generalizes to the Reissner-Nordstr{\"o}m (RN) geometry. Besides the event horizon, this geometry also comes with an (inner) Cauchy horizon (CH). As first suggested by Penrose \cite{penrose68} and subsequently verified in \cite{Poisson:1989zz,Poisson:1990eh,Ori:1991zz}, the Cauchy horizon appearing in the RN solution is unstable and suffers from a mass-inflation effect. In short, the cross-flow of ingoing and outgoing matter leads to an eternal, exponential growth of the mass function at the Cauchy horizon.

This instability for the RN black hole, and other geometries containing time-like singularities, also persists at the level of quantum field theory in curved spacetime. Studying the growths of the expectation value of the stress energy tensor of a scalar field at the Cauchy horizon, it was shown that quantum effects do not cure the instability \cite{Casals:2016odj, Casals:2018ohr, Sela:2018xko, Casals:2019jfo, Zilberman:2019buh, Klein:2021ctt,  Zilberman:2022aum, Klein:2024sdd, Arrechea:2024ajt}. This conclusion also persists when the RN black hole is embedded into de Sitter space \cite{Hollands:2019whz, Hollands:2020qpe}. So quantum effects do not tame the mass-inflation instability in these cases.  

The breakdown of the laws of physics at a spacetime singularity motivates the construction of regular black hole geometries which resolve this singularity in one way or another. Starting from the works by Bardeen \cite{BardeenBH, Ayon-Beato:2000mjt}, there is now a rich literature on this subject \cite{Bambi:2023try,Lan:2023cvz,Carballo-Rubio:2025fnc}, including the proposals by Dymnikova \cite{Dymnikova:1992ux, Dymnikova:2001fb, Platania:2019kyx}, Bonanno-Reuter  \cite{Bonanno:2000ep, Koch:2014cqa}, Hayward \cite{Hayward:2005gi}, Simpsson-Visser \cite{Simpson:2019mud}, and lately also the regular geometries obtained from a model of gravitational collapse within the gravitational asymptotic safety program \cite{Bonanno:2023rzk}. Moreover, there is currently a significant effort in obtaining such solutions from dynamical principles, e.g., by resorting to non-linear electrodynamics \cite{Ayon-Beato:1998hmi,Bronnikov:2022ofk, Malafarina:2022oka} or studying infinite towers of curvature monomials \cite{Bueno:2024dgm}. A feature frequently encountered in these regular geometries is that they also possess a Cauchy horizon, making the geometry similar to the (RN) solution.\footnote{Lately, this idea has spurred the development of regular black hole geometries where the CH is absent \cite{Biasi:2022ktq, Ovalle:2024wtv, Casadio:2025pun, Calza:2025mrt, Hale:2025ezt}.} On this basis, it was believed for a long time that also regular black holes exhibiting a Cauchy horizon would suffer from the mass-inflation instability, rendering the Cauchy horizon singular \cite{Carballo-Rubio:2018pmi}.

Returning to the classical analysis, the study of the Ori model \cite{Ori:1991zz} (also see \cite{Hamilton:2008zz,Bonanno:2022rvo} for reviews) for regular black hole geometries revealed a remarkable feature: in some cases, including the Hayward geometry and the Bonanno-Reuter black holes, the exponential growth of the mass function is replaced by a power-law scaling \cite{Bonanno:2020fgp,Bonanno:2023qhp}. The exponent appearing in the power-law is further reduced once Hawking radiation is incorporated in the analysis \cite{Bonanno:2022jjp}. In these cases, mass inflation may reduce to a transient phenomenon \cite{Carballo-Rubio:2024dca}. Moreover, the character of the singularity building up at the Cauchy horizon may change from a strong to a weak singularity \cite{Bonanno:2020fgp, Bonanno:2022jjp} according to the definitions given in \cite{Tipler:1977zza, Krolak:1986pno, Krolak:1987, Ori:1991zz, Bonanno:1994qh, Burko:1999zv}. Ultimately, these drastic deviations from the RN analysis can be traced back to the fact that equation determining  the dynamics of the mass function is no longer linear but quadratic in the mass. This leads to new late-time behaviors which are absent in the RN analysis.

The present work expands on this picture in two ways. Building on the Ori-model \cite{Ori:1991zz}, we present a complete phase space analysis for the regular black holes of Bardeen, Hayward, Dymnikova, and AS-collapse. In all cases, we identify the quasi-fixed points governing the late-time dynamics of the models. For the Hayward geometry, this reveals that the conversion of the mass-inflation effect to a transient phenomenon is generic. The power-law attractor controlling the late-time dynamics is reached from generic initial conditions. The study of the Dymnikova \cite{Dymnikova:1992ux} and AS-collapse model \cite{Bonanno:2023rzk} extends the systematic study to regular black holes whose mass functions have non-analytic properties. For the Dymnikova case, it is shown that the phase space has a rich structure in which eternal mass inflation is rather generic. In contrast, the growth of the mass function obtained from the AS-collapse model is weaker than power law and only growths logarithmically. Moreover, no appreciable transient mass-inflation episode occurs before the attractor is reached. The strength of the singularity building up in this case is then even weaker than the one found for the Hayward case. 

The rest of this work is organized as follows. Sect.\ \ref{Sect.1a} introduces the geometries investigated in this work and reviews the ``classical'' mass-inflation instability for the RN geometry. Subsequently, we introduce the Ori model in Sect.\ \ref{Sect.2} and discuss its the generic late-time dynamics resulting from attractive quasi-fixed points in Sect.\ \ref{Sect.2.2}. The dynamics for selected regular black holes is covered in Sect.\ \ref{Sect.3} and we close with our conclusions and a brief outlook in Sect.\ \ref{Sect.4}. Our definitions of curvature invariants are collected in App.\ \ref{App.A} and the technical details underlying the analytic analysis of the AS-collapse model are given in App.\ \ref{App.D}. Throughout this work, we set Newton's constant $G=1$, so that all dimensionful quantities are measured in Planck units.

\section{Cauchy horizon instability}
\label{Sect.1a}
%
\begin{figure}[t!]
\begin{center}
\includegraphics[width=0.42 \textwidth]{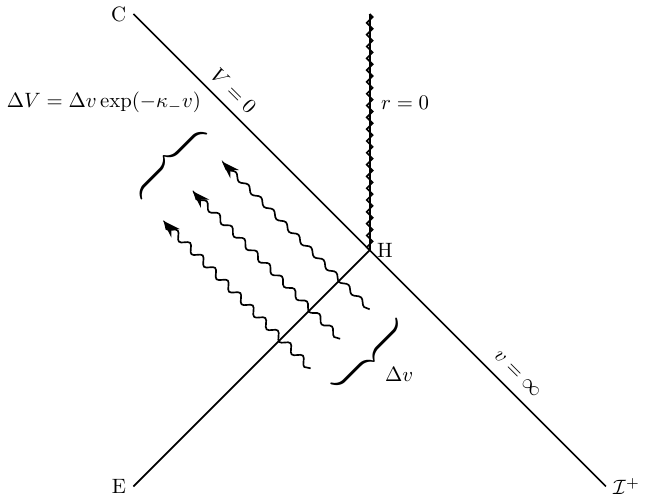}
\end{center}
\caption{\label{Fig.1a} Radiation falling in to a Reissner-Nordstr{\"o}m black hole. The light is emitted during a period of external advanced time $\Delta v$ and a observer inside the hole receives the signals during an interval of internal advanced time $\Delta V$.}
\end{figure}

We are interested in spherically symmetric black hole geometries. In this case, one can always find coordinates such that the line element takes the form
\begin{equation}
\label{eq.line.x}
ds^2 = g_{ab} dx^a dx^b + r^2 d\Omega^2 \, , \quad a,b = 0,1 \, .
\end{equation}
Here $d\Omega^2 = d\theta^2+\sin^2 \theta \, d\phi^2$ and $r=r(x^a)$ sets the radius of the $2$-sphere at the point $x^a$. We then introduce the Misner-Sharp mass $M(x^a)$ via the gradient
\be\label{def.Mx}
f(x^a) \equiv g^{ab} \, \p_a r \, \p_b r = 1 - \frac{2M(x^a)}{r} \, . 
\ee
Static geometries are then completely fixed by the radial profile $M(r)$. In particular, the Schwarzschild metric has $M(r) = m$ with $m$ being the ADM-mass of the geometry. The examples studied in this work have\footnote{In \eqref{geo.log}, with respect to \cite{Bonanno:2023rzk}, we implemented an analytic continuation of $M(r)$ so that the Misner-Sharp mass remains real for all values $m$.}
\begin{subequations}\label{eq:geometries}
\begin{align} \label{geo.RN}
&\mathrm{RN:} & \; M(r) = & \, m - \frac{e^2}{2r} \, ,  \\ \label{geo.Bardeen}
&\mathrm{Bardeen:} & \; M(r) = & \, \frac{m r^3}{(r^2 + a^2)^{3/2}} \, ,  \\ \label{geo.Hayward}
&\mathrm{Hayward:} & \; M(r) = & \, \frac{m r^3}{r^3 + 2 m l^2} \, , \\
\label{geo.Dymnikova}
& \mathrm{Dymnikova:} & \; M(r) = & \, m \left[ 1 - \exp\left( - \frac{r^3}{\gamma^2 m} \right) \right] \, , \\ \label{geo.log}
&\text{AS-collapse:} & \; M(r) = & \, \frac{r^3}{12 \xi} \log\left( 1 + \frac{6 \xi m}{r^3} \right)^2  \, . 
\end{align}
\end{subequations}
Here $m$ is the mass of the black hole which can be measured by using Kepler's third law outside the hole. In the RN case, $e$ is its the electric charge of the black hole. In the geometries \eqref{geo.Bardeen}-\eqref{geo.log} the parameters $\{a,l,\gamma, \xi\}$ encode the deviation of the geometry from the Schwarzschild solution. The regular black hole geometries \eqref{geo.Bardeen}-\eqref{geo.Dymnikova} replace the spacetime singularity of the Schwarzschild black hole by a regular de Sitter core. This is readily seen by expanding $M(r)$ for small values of $r$, yielding $M(r) \simeq c r^3$ with the model-dependent constant $c>0$. These geometries realize the limiting curvature as mechanism of singularity avoidance which has an old and prominent role in the literature \cite{1982ZhPmR..36..214M, Markov:1984ii, Frolov:1989pf, Frolov:1988vj, Frolov:2016pav}. In contrast, the AS-collapse geometry replaces the singularity by a core of matter of finite size whose boundary is located within the Cauchy horizon \cite{Bonanno:2023rzk}.

Our focus is on the case where the equation $f(r) = 0$ has two solutions $r_{+} > r_- > 0$ which correspond to the position of the (outer) event horizon and the (inner) Cauchy horizon, respectively. We then define the surface gravity $\kappa_\pm$ of the horizons 
\be\label{def.surface}
\kappa_\pm  \equiv \pm \frac{1}{2}\frac{\partial f(r)}{\partial r} \bigg|_{r=r_{\pm}} \, ,
\ee
with $\kappa_\pm > 0$ by definition.

We outline the fundamental physical mechanism responsible for the CH instability of the RN geometry \eqref{geo.RN}. Adopting Schwarzschild coordinates, the line element is
\begin{equation}
    ds^2=-f(r) dt^2+\frac{1}{f(r)}dr^2+d\Omega^2 \, ,
\end{equation}
with the function $f(r)$ given by
\begin{equation}
    f(r) = 1- \frac{2 m}{r}+\frac{ e^2}{r^2} \, . 
\end{equation}
The roots solving $f(r)=0$ are
\begin{equation}
    r_{\pm}=m \pm \sqrt{m^2+e^2}
\end{equation}
and correspond to the event horizon at $r=r_+$ and the Cauchy horizon at $r=r_-$. The surface gravity \eqref{def.surface} evaluates to 
\begin{equation}
    \kappa_{\pm}=\frac{r_+-r_-}{2 \, r^2_\pm} \, .
\end{equation}

The tortoise coordinate $r_\ast$, defined by 
\begin{equation}
    r_\ast=\int \frac{dr}{f(r)} \, , 
\end{equation}
is used to define the Eddington-Finkelstein double null coordinates 
\begin{equation}
    \bar v = t+r_\ast,\quad \bar u = t-r_\ast \, . 
\end{equation}
These reduce to the usual advanced and retarded times far from the black hole. These coordinates are good in the exterior of the black hole where the line element reads
\begin{equation}
    ds^2=- f \, d\bar{u} \, d \bar{v} + r^2 d\Omega^2 \, .
\end{equation}
 At the event horizon there is a coordinate singularity where $\bar u \rightarrow \infty $. A free-falling observer will use a retarded time 
\begin{equation}
    U_{obs}= e^{-\kappa_+ \bar{u}} \, ,
\end{equation}
and will not notice anything special when crossing the event horizon. Observers far from the black hole will measure the outgoing signals emitted by our observer to be infinitely red shifted when the latter approaches the event horizon. 

In order to extend the RN solution to the inside of the event horizon, we need to use a different coordinate patch,
\begin{equation}
    v= t+r_\ast,\quad u =r_\ast-t
\end{equation}
where $u \in [-\infty,\infty]$ as $r \in [r_-,r_+]$. The metric in the interior is
\begin{equation}
    ds^2 = f dudv +r^2 d\Omega^2 \, . 
\end{equation}
The coordinate singularities at the horizons can now be removed via a ``Kruscalization" of the $u,v$ couple. For the CH one defines Kruskal coordinates
\begin{equation}\label{def.Kruskal}
    \kappa_- U = - e^{-\kappa_- u}, \quad \kappa_- V = -e^{-\kappa_-v} \, . 
\end{equation}
Near the CH  we have $f\simeq-2 e^{-\kappa_-(u+v)}$ and the line element reads
\begin{equation}
    ds^2=-2dUdV + r^2 d\Omega^2 \, .
\end{equation}

Let us now consider a pencil of radiation entering the black hole as in Fig.\ \ref{Fig.1a}. Assume its duration is finite and given by $\Delta v$ as measured by an external observer. The local observers located near the CH use instead the Kruskal coordinates to measure the duration of the influx $\Delta V$. From the relations \eqref{def.Kruskal}, we infer 
\begin{equation}
\Delta V = -e^{-\kappa_- \, v}  \, \Delta v \, . 
\end{equation}
When the signal reaches the CH at $v\rightarrow \infty$, there is an infinite blueshift of the radiation for the local observer. The fact that the CH is therefore a surface of infinite blueshift has led Penrose to the conjecture that the spacetime region beyond the CH is unphysical. 

The first attempt to take into account the backreaction of the perturbations at the CH has been discussed by Poisson and Israel \cite{Poisson:1989zz}.   In a realistic collapse the CH is subject to the cross flow of gravitational waves produced by the collapsing star and by the flux of gravitational radiation which is partially reflected and partially transmitted by the internal potential barrier. In the model of Poisson and Israel this effect is modeled by a stress tensor for a null cross flow of radiation
\begin{equation}
\label{pit}
    T_{\alpha\beta}=\frac{L_{in}(V)}{ 4 \pi r^2}\partial_\alpha V\partial_\beta V+
    \frac{L_{out}(U)}{4 \pi r^2} \partial_\alpha U \partial_\beta U \, ,
\end{equation}
where the luminosity function $L_{in}$, in the Kruskal coordinate $V$, has the following form
\be
\begin{aligned}
    L_{in}(V)&=\frac{d m_{in}}{dv}\left ( \frac{dv}{d V} \right )^2 \\
    &=\frac{\beta}{(-\kappa_{-}V)^2} [-\log (-\kappa_ {-}V)]^{-p} \, .
\end{aligned}
\ee
When the CH is approached, in the limit $V\rightarrow 0^{-}$, the luminosity function $L_{in}(V)$ diverges. As a consequence the mass function and the local curvature invariant $\Psi_2$, defined in eq.\ \eqref{def:psi2}, diverge as $1/V \propto - \, e^{\kappa_- v}$ and a null, weak singularity forms at the CH \cite{Poisson:1989zz,Poisson:1990eh}.

\section{Ori Model}
\label{Sect.2}
The Ori model \cite{Ori:1991zz} is an analytical model that allows to  discuss the backreaction of the geometry to the instability of the CH. It is based on the Poisson-Israel model and approximates the outgoing continuous flux in (\ref{pit}) as a thin localized shell, $L_{out} = \delta (U-U_0)$. This section introduces the relevant features of the model.

\begin{figure}[t!]
\begin{center}
\includegraphics[width=0.42 \textwidth]{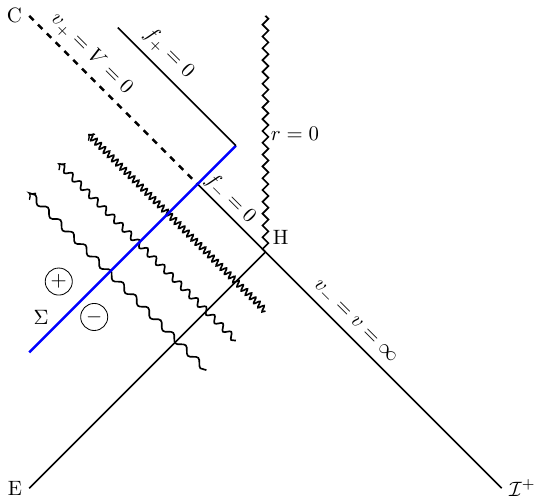}
\end{center}
\caption{\label{Fig.1} Penrose diagram illustrating the Ori-model of the interior for a Reissner-Nordstr{\"o}m black hole \cite{Ori:1991zz}. The event horizon (E) and Cauchy horizon (C) are marked with a straight and dashed line, respecively. 
Outgoing radiation is modelled by a thin shell $\Sigma$ which crosses the Cauchy horizon. A singularity forms in the $+$ region on the Cauchy horizon. It is represented by a bold-dashed line as it is much weaker than the $r=0$ singularity.}
\end{figure}

We start from the static geometry and introduce perturbations in the form of null-crossflowing radiation \cite{Poisson:1989zz,Poisson:1990eh}, see Fig.\ \ref{Fig.1} for illustration. Following the simplifications introduced in \cite{Ori:1991zz}, we model the outgoing energy flux by a spherically symmetric, infinitesimally thin, pressureless null shell placed between the event and the Cauchy horizon. The outgoing perturbation is given by an infinitesimally thin shell $\Sigma$. This shell divides the spacetime into the region $\cM_+$ inside the shell and the region $\cM_-$ outside of $\Sigma$. In each region, we choose ingoing Eddington-Finkelstein coordinates
\begin{equation}
\label{geo.region}
ds^2_{\pm}=-f_{\pm}(v_{\pm}, r)dv^2_{\pm}+2drdv_{\pm}+r^2d\Omega^2 \, . 
\end{equation}
Here we use subscripts $\pm$ for coordinates on $\cM_\pm$ and anticipated that the radial coordinate can be taken the same in both regions. The relation between $v_\pm$ is fixed by noting that $\Sigma$ moves lightlike, i.e.\ we have 
\begin{equation}
\label{vpm.relation}
f_- dv_- = 2 dr = f_+ dv_+ \, . 
\end{equation}
We use this relation to express $v_+$ in terms of the coordinate on $\cM_-$, $v \equiv v_-$. Furthermore, \eqref{vpm.relation} implies that the position of the shell evolves according to
\be\label{eq.Rdiv}
\frac{\p R(v)}{\p v} = \left. \frac{1}{2} \,  f_-(v,r) \right|_\Sigma \, . 
\ee

Subsequently, we have to relate the Misner-Sharp mass $M_\pm$ in the two sectors. This relation follows from the junction conditions imposed at the position of $\Sigma$. Following \cite{Ori:1991zz}, we construct these conditions based on the dynamics given by general relativity. Writing Einstein's equations in each sector and expressing the result in terms of $M(r,v)$ yields
\be\label{eq.Einstein}
\frac{\partial M(v, r)}{\partial r}=-4\pi r^2 T_{v}^{\, \,v} \, , \quad \frac{\partial M(v, r)}{\partial v}=4\pi r^2 T_{v}^{\, \,r} \, ,
\ee 
with $T_{rr} = 0$. Continuity across $\Sigma$ requires \cite{Barrabes:1991ng}
\be \label{eq.junction}
\left[ T_{\mu\nu} s^\mu s^\nu \right] = 0 \, .
\ee
Here the null generators can be taken as $s_\pm^\mu = (2/f_\pm , 1,0,0)$ and eq.\ \eqref{eq.junction} implies that the scalar quantity in the square brackets should be continuous across the shell. Recasting \eqref{eq.junction} in terms of the Misner-Sharp mass then gives the desired relation between the mass functions in the two sectors
\be\label{eq.junction.eq}
\left. \frac{1}{f_+^2} \, \frac{\p M_+}{\p v_+} \right|_\Sigma = \left. \frac{1}{f_-^2} \, \frac{\p M_-}{\p v_-} \right|_\Sigma \, . 
\ee 
The relation \eqref{vpm.relation} allows to recast this expression in terms of $v$
\be\label{eq.junction.2}
\left. \frac{1}{f_{+}(v, r)}\frac{\partial M_{+}(v, r)}{\partial v} \right|_{\Sigma}=F(v)
\ee
where we defined
\be\label{def.Fv}
F(v)\equiv \left. \frac{1}{f_{-}(v, r)}\frac{\partial M_{-}(v, r)}{\partial v} \right|_{\Sigma} \, .
\ee
Taking $M_+$ as a function of $r$ and $m_+(v)$ (with $r$ and $v$ being independent variables), this equation can be written as
\begin{equation}\label{eq.mp.1}
\dot{m}_+ = \; \left. \left[ \frac{\p M_+(r,m_+)}{\p m_+} \right] ^{-1}    \, f_+(r,m_+)  \, F(v) \right|_\Sigma \, .   
\end{equation}

Eqs.\ \eqref{eq.Rdiv} and \eqref{eq.mp.1} specify the dynamics of the model \emph{independently of the specific background geometry}. The generic expressions are readily adapted to the specific cases introduced in eq.\ \eqref{eq:geometries}. Their derivation relies on properties of the shell and the junction condition derived from general relativity only. 

The completion of the dynamics requires specifying the perturbation of the mass function in the exterior of the shell. Here we assume that the late-time behavior of the perturbation follows Price's law \cite{Price:1971fb, Price:1972pw}
\be\label{def.Price}
\mminus(v)=m_{0}-\frac{\beta}{(v/v_{0})^{p-1}}\, ,  \quad p \geq 12\, .
\ee
Here, $m_0$ is the asymptotic mass in the region $\cM_-$, $\beta$ is a parameter with the dimension of a mass and $v_{0}$ is the initial value of $v$ which is set to one in the sequel. The value $p=12$ corresponds to the power-law decay of the radiative tail associated with quadrupole waves.

\section{Universal late-time dynamics}
\label{Sect.2.2}
Starting from eqs.\ \eqref{eq.Rdiv} and \eqref{eq.junction.2}, we now discuss universal features of the dynamics encoded in the Ori model. In order to lighten our notation, derivatives with respect to $v$ and $r$ are denoted by a dot and a prime, respectively. 

We start with eq.\ \eqref{eq.Rdiv}. Once Price's law \eqref{def.Price} is adopted, this turns into a first order differential equation for the position of the shell $R(v)$. Obviously, $R(v) = r_-$ is a solution to this equation, since in this case the right-hand side (RHS) vanishes due to the horizon condition $f(r_-)=0$. The dynamics of the shell impacting on the CH can then be studied by adopting the power-law ansatz
\be\label{AnsatzR}
R(v) = r_{-}+\frac{1}{v^s}\sum_{n=0}^{\infty} \, \frac{a_n}{v^n} \, . 
\ee
Here $s$ is a to-be-determined parameter and the coefficients $a_n$ encode the information on how the shell approaches the horizon. We now demonstrate universality in \eqref{AnsatzR}, showing that it is largely independent of the specific form of $f(r)$.

In order to highlight the structure of eq.\ \eqref{eq.Rdiv}, we recast the perturbations as
\be\label{eps.expansion}
m_-(v) = m_0 + \epsilon_1 \, , \qquad R(v) = r_- + \epsilon_2 \, . 
\ee
The explicit form of $\epsilon_1$ and $\epsilon_2$ can be extracted from eqs.\ \eqref{def.Price} and \eqref{AnsatzR} and read
\be\label{eps.def}
\epsilon_1 = -\frac{\beta}{v^{p-1}} \, , \quad  
\epsilon_2 = \frac{1}{v^s}\sum_{n=0}^{\infty} \, \frac{a_n}{v^n} \, .   
\ee
Next, we substitute \eqref{eps.expansion} into \eqref{eq.Rdiv} and expand to first order in $\epsilon_i$:
\be\label{Rvepsexp1}
\begin{split}
2 \, \dot{\epsilon}_2 = & f_-(r_-, m_0) \\ &   + \frac{\p f_-( r_-, m_0)}{\p m_0} \, \epsilon_1 +  \frac{\p f_-(r_-, m_0)}{\p r_-} \, \epsilon_2  \\ & \, + \mathcal{O}(\epsilon^2) \, . 
\end{split}
\ee
The first term on the RHS vanishes due to the horizon condition. Moreover, the prefactor multiplying $\epsilon_2$ is proportional to the surface gravity \eqref{def.surface}. Encoding the model-dependent information from the first term by defining
\be\label{def.betat}
\tilde{\beta} \equiv \frac{1}{2} \, \frac{\p f_-(r_-, m_0)}{\p m_0} \, , 
\ee
eq.\ \eqref{Rvepsexp1} takes the form
\be\label{Rvepsexp2}
\dot{\epsilon}_2 = \tilde{\beta} \, \epsilon_1 - \kappa_- \, \epsilon_2 + \mathcal{O}(\epsilon^2) \, . 
\ee
This equation should be read as an expansion in powers of $1/v$. The parameter $s$ and the coefficients $a_n$ are then found using the Frobenius method. In a non-trivial solution, the leading powers of $1/v$ are provided by the terms proportional to $\epsilon_1$ and $\epsilon_2$. This fixes the free parameter $s$ in eq.\ \eqref{AnsatzR} to
\be\label{Rscaling}
s = p-1 \, . 
\ee
Thus the leading term in $\epsilon_2$ follows the same power law as Price's law, \emph{as long as the horizon comes with a non-vanishing surface gravity}. This result is independent of the underlying geometry.

The coefficients $a_n$ are then readily found by solving eq.\ \eqref{Rvepsexp2} order-by-order in $1/v$. The first few coefficients are
\begin{subequations}\label{Rvdynamics}
\begin{align}\label{a0coeff}
a_0 = & \, - \frac{\beta \, \tilde{\beta}}{\kappa_-} \, , \\ \label{ancoeff}
a_{n+1} =  & \, \frac{1}{\kappa_-} \left( p-1+n \right) \, a_n \, , \quad n < p -2 \, . 
\end{align}
\end{subequations}
The recursion relation \eqref{ancoeff} shows that the time-scale associated with the perturbation settling on the CH is set by the surface gravity $\kappa_-$. For $n \ge p - 1$ onward, terms quadratic in $\epsilon$ start contributing and modifying the linear relation \eqref{ancoeff}. While it is possible to work out these contributions systematically, this is not needed when discussing the late-time behavior of the model analytically. Moreover, Price's law contains the leading fall-off behavior and subleading terms (associated with higher-order multipole moments in the perturbations) may modify $\epsilon_1$ at subleading orders. Therefore, only the lowest order terms in \eqref{Rvdynamics} are robust with respect to these additional corrections.

In the next step, we focus on the late-time behavior entailed by eq.\ \eqref{eq.mp.1}. In order to aid the discussion, we introduce the abbreviation
\begin{equation}\label{F.eq2}
g(r,m_+) \equiv \left. \left[ \frac{\p M_+(r,m_+)}{\p m_+} \right]^{-1} \right|_\Sigma \, . 
\end{equation}

The expression \eqref{eq.mp.1} allows to distinguish several cases. First, the RHS can be linear in $m_+$. This happens if the Misner-Sharp mass is a linear function of the asymptotic mass and applies to the RN black hole \eqref{geo.RN} and the regular Bardeen black hole \eqref{geo.Bardeen}. This leads to the standard mass-inflation instability where the Misner-Sharp mass grows exponentially in $v$, c.f.\ eq.\ \eqref{mass-inflation-instability}.
Second, the RHS may have non-trivial roots $m_{+,*}$ where
\begin{equation}\label{def.attractor}
g(r_-,m_{+,*}) = 0 \, . 
\end{equation}
Clearly, $m_{+,*}$ is a solution, if \eqref{eq.mp.1} is evaluated at the Cauchy horizon. This gives rise to quasi-fixed points in the dynamics. These can either be attractive or repulsive. Notably, it is an attractive quasi-fixed point that underlies the late-time attractors discovered in \cite{Bonanno:2020fgp}.

We proceed by determining the generic late-time behavior induced by these attractors, indicating a quantity that is evaluated at the attractor by $*$. Furthermore, we make the generic assumptions that, first, the metric functions are continuous and sufficiently differentiable everywhere, second, that $f_+(r_-, m_{+,*})$ takes a finite and non-vanishing value, and, third, that
\begin{equation}\label{F.eq3}
\lim_{v \rightarrow \infty} F(v) \equiv F_*
\end{equation}
approaches a finite value asymptotically. Finally, we assume that the late-time dynamics of $m_+(v)$ can be captured by the Frobenius ansatz
\begin{equation}\label{eq.mp.Frobenius}
m_+(v) = m_* + \frac{1}{v^t} \sum_{n=0} \frac{b_n}{v^{n}} \, , 
\end{equation} 
where $t$ and $b_n$ are coefficients to be determined. Following the notation introduced in eq.\ \eqref{Rvepsexp2}, we denote the perturbations around the attractor by
\begin{equation}\label{def.eps3}
\epsilon_3 \equiv \frac{1}{v^t} \sum_{n=0} \frac{b_n}{v^{n}}  \, . 
\end{equation}

We proceed by expanding $g$ around the attractor
\begin{equation}\label{F.eq5}
\begin{split}
& g[R(v),m_+(v)]  \simeq g(r_-,m_*) \\ &  \quad + \left( \frac{\p g}{\p r}\right)_* \, \epsilon_2 + \left( \frac{\p g}{\p m_+}\right)_* \epsilon_3  + \mathcal{O}(\epsilon^2) \, . 
\end{split}
\end{equation}
The condition \eqref{def.attractor} ensures that the first term in the expansion vanishes. Eq.\ \eqref{eq.mp.1} then states that the late-time behavior of the system is captured by
\begin{equation}\label{eq.mp.asymp}
\dot{\epsilon}_3 = \left[ \left( \frac{\p g}{\p r}\right)_* \, \epsilon_2 + \left( \frac{\p g}{\p m_+}\right)_* \epsilon_3 \right] \, f_{+,*} \, F_* \, . 
\end{equation}
Remarkably, the subleading terms in $f_+$ and $F(v)$ do not play a role. The fall-off of the LHS is then subleading to the fall-off of the RHS. Thus, for a consistent solution, the leading terms in the straight bracket have to vanish. This fixes
\begin{equation}\label{eq.tasymp}
t = p - 1
\end{equation}
and
\begin{equation}\label{eq.b0}
b_0 = - \left( \frac{\p g}{\p m_+} \right)_*^{-1} \, \left( \frac{\p g}{\p r}\right)_* \, a_0 \, .
\end{equation}
The coefficient $b_1$ is then fixed from the order $1/v^{p}$ in the expansion
\begin{equation}\label{F.eq8}
b_1 = - \left[  \frac{(p-1) \, b_0}{f_{+,*} \, F_*} + \left( \frac{\p g}{\p r}\right)_* \, a_1 \right] \, \left( \frac{\p g}{\p m_+} \right)_*^{-1} \, . 
\end{equation} 
Notably, subleading corrections to $F_*$ and $f_{+,*}$ do not enter into $b_1$. These are needed to determine $b_2$ only. The reason is that these terms are multiplied by the equation determining $b_0$ and therefore drop out of \eqref{F.eq8}.

Based on this analysis, we arrive at the following conclusion.  Eq.\ \eqref{F.eq8} is inhomogeneous in the sense that it contains extra contributions which are not generated by the expansion of $g$. As a consequence
    \begin{equation}\label{F.eq9}
    g \simeq - \frac{p-1}{f_{+,*} \ F_*} \frac{b_0}{v^{p}} \, . 
   \end{equation}
    This scaling behavior is dictated by the structure of eq.\ \eqref{eq.mp.1} and hinges on very few generic assumptions only. Notably, the factor $g$ also appears in $M_+$. In this way, the result \eqref{F.eq9} sets the late-time behavior of the Misner-Sharp mass evaluated at the position of the shell, 
    \be\label{misner-sharp-v}
    M_+(v) = M_+[R(v), m_+(v)] \, , 
    \ee
    when approaching the attractor.

At this point, we established that the scaling of $g$ related to the late-time attractor $m_{+,*}$ is rather generic. A way to avoid this conclusion is to violate one of the assumptions underlying its derivation. For example, the AS-collapse model \eqref{geo.log} and the Dymnikova black hole \eqref{geo.Dymnikova} violates the fact that $f_{+,*}$ is finite. In addition, the Dymnikova geometry gives rise to a discontinuity as $m \rightarrow 0$ either from above or below. This suggests that these models could give rise to new scaling behaviors which differ from the one induced by \eqref{F.eq9}. This will be investigated on a case-by-case basis in the next section.

\section{Regular black holes (in)stability}
\label{Sect.3}
Upon discussing general features of the dynamics encoded in the Ori model, we illustrate these features for the specific examples introduced in eq.\ \eqref{eq:geometries}. We start by analyzing the phase space of the solutions in Sect.\ \ref{Sect.3.3}. Analytic results on the late-time attractor are collected in Sect.\ \ref{Sect.3.1} and we verify these features at the level of numerical solutions in Sect.\ \ref{Sect.3.2}. Sect.\ \ref{sect.6} then briefly comments on the strength of the singularity building up at the CH. The results of this section are conveniently summarized in Table \ref{Tab.summary_relevant}.
\subsection{Phase Space}
\label{Sect.3.3}
\begin{figure*}[t]
\includegraphics[width = 0.48\textwidth]{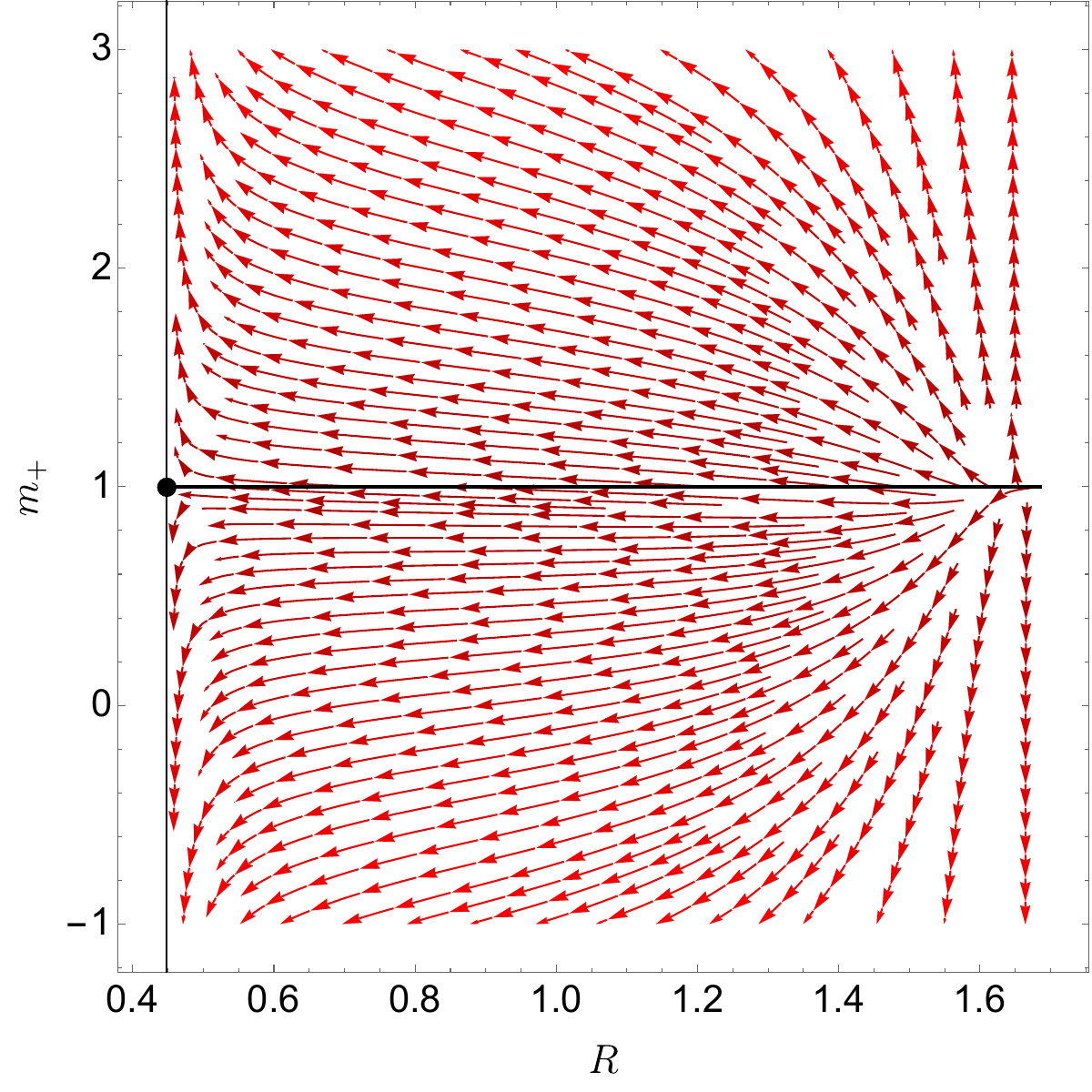} \, 
\includegraphics[width = 0.48\textwidth]{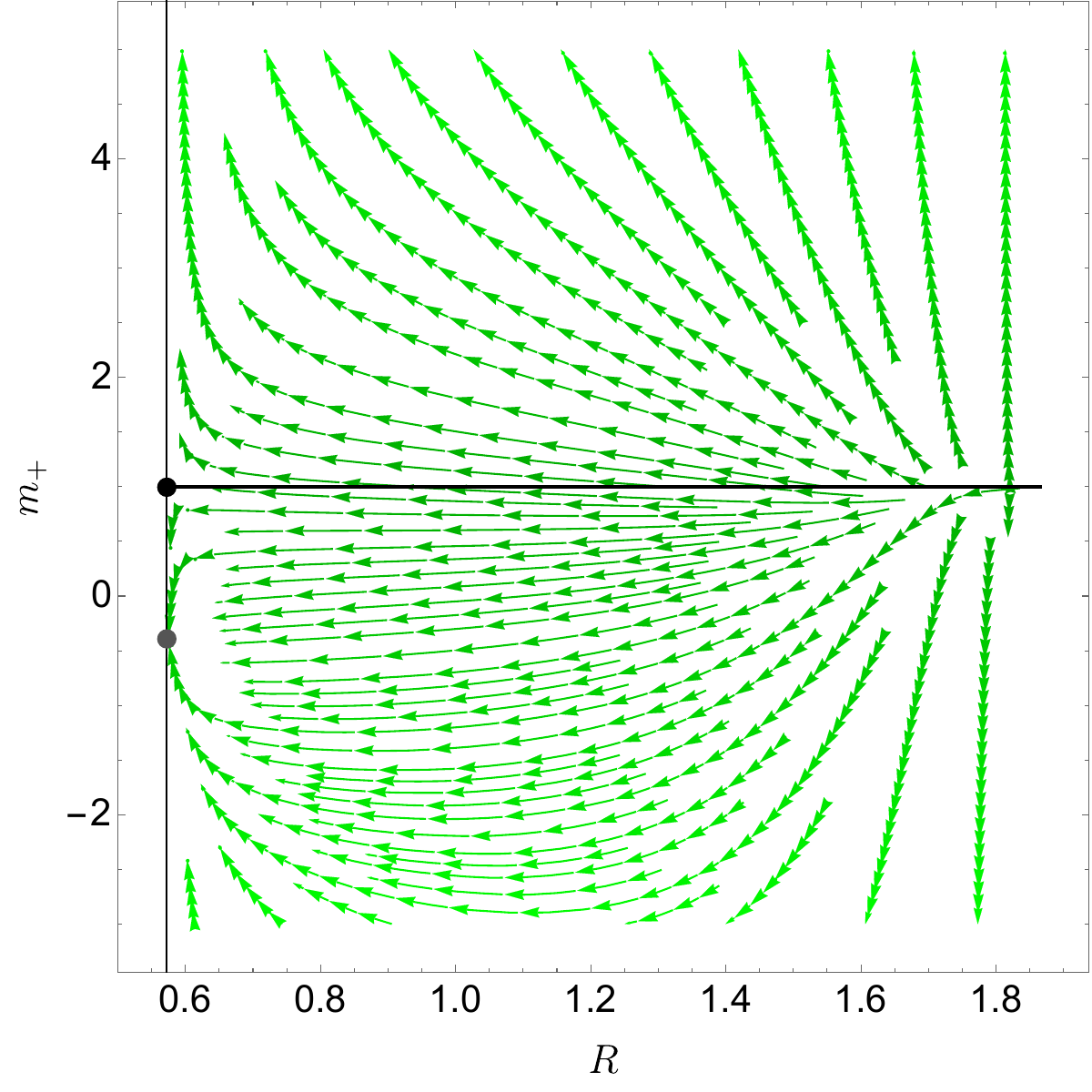} \\
\includegraphics[width = 0.48\textwidth]{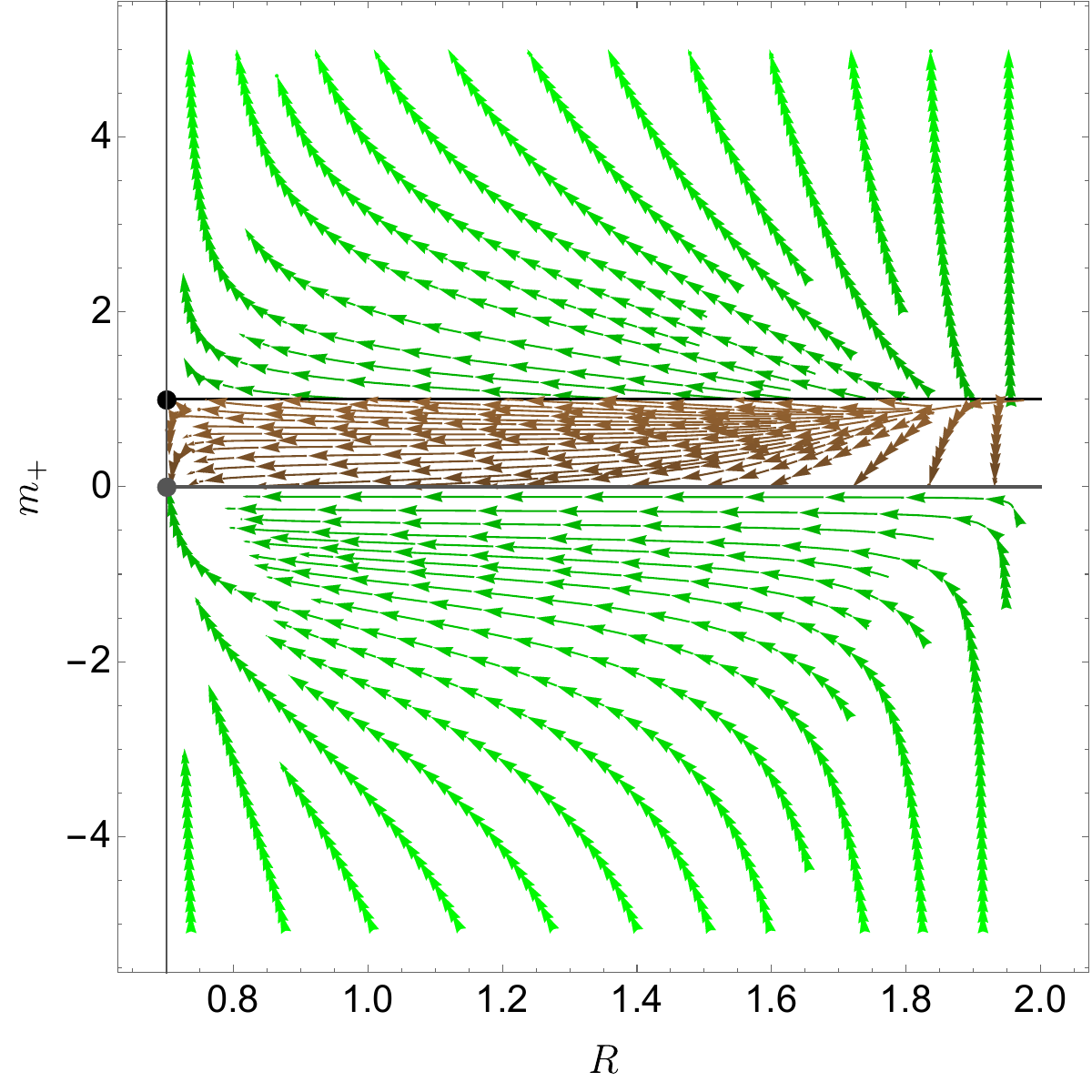} \, 
\includegraphics[width = 0.48\textwidth]{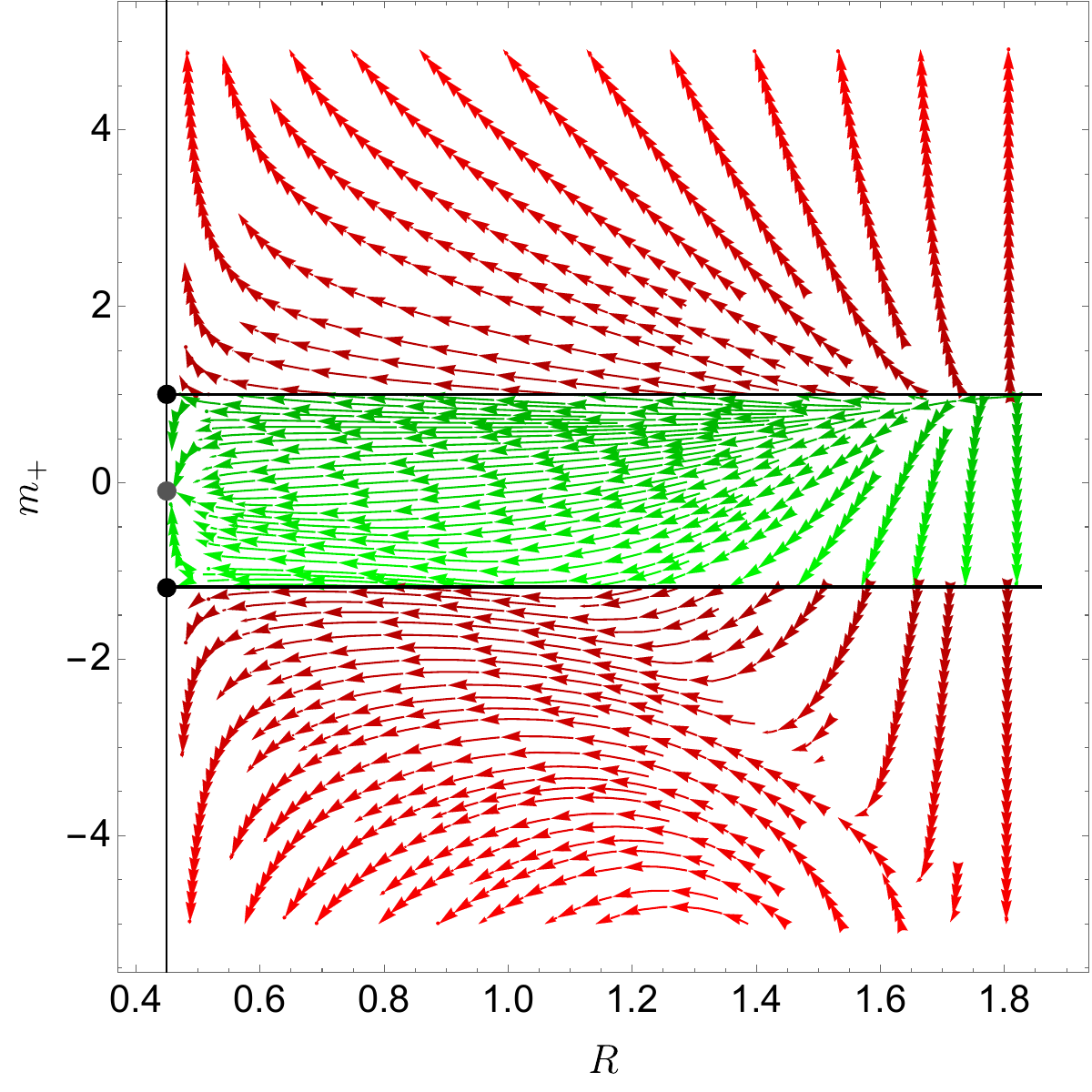} \\
\caption{\label{Fig.phasespace} Illustration of the dynamics arising from the Ori model for the Bardeen geometry (top left), the Hayward geometry (top right), the Dymnikova geometry (bottom left), and the AS-collapse (bottom right). The CH is located at the black vertical line. The position of the attractive (repulsive) quasi-fixed points $m_{+,*}$ are marked with gray (black) dots and are extended by horizontal lines. The stream lines are indicative for the phases encountered in the models where flows attracted to (repelled by) a quasi-fixed point are indicated in green (red). In addition, the Dymnikova geometry gives rise to an additional phase (brown) where the solutions terminate at the discontinuity $m_+ = 0$  at a finite value $v$.}
\end{figure*}

As we demonstrated in Sect.\ \ref{Sect.2.2}, the impact of the shell on the CH is independent of the specific features of the geometry. Differences in the stability of the horizon arise from the dynamics of $m_+(v)$. In the following, we discuss the key properties of these equations based on the phase space diagrams shown in Fig.\ \ref{Fig.phasespace}. The parameters used in each geometry are listed in Table \ref{tab:initflow}. In particular, we have fixed the mass in Price's law to $m_0 = 1$ and adjusted the free parameters in such a way that the surface gravity $\kappa_-$, setting the time-scale of the process, is $\kappa_-=1$ in all cases. In all cases we highlight the values $m_{+,*}$ of the attractive (repulsive) quasi-fixed points as gray (black) horizontal lines. The stream plots are then generated at a fixed instance of time $v_{\rm section}$. This entails that the streams could actually cross these lines. Throughout the discussion, we color-code streams repelled by a quasi-fixed point in red, trajectories ultimately ending at an attractive quasi-fixed point in green, and solutions terminating at a finite value $v$ in brown. \\

\begin{table*}[htbp!]
\begin{ruledtabular}
\renewcommand{\arraystretch}{1.5}
\begin{tabular}{l|cccc|ccc|ccc|c}
\textit{Solution}    & \textit{Parameter} & $r_-$ & $r_+$ & $\kappa_-$ & $m_{+,*}^{\rm rep}$ &  $m_{+,*}^{\rm att}$ & $m_{+,*}^{\rm rep}$ & $m_0$ & $\beta$ & $p$ & $v_{\rm section}$  \\ \hline
     Bardeen    & $a=0.586$ & $0.447$ & $1.686$ & $1$ & $1$ & $-$ & $-$ & $1$ & $1$ & $12$ & $1.7$\\
     Hayward & $l=0.483$ & $0.571$ & $1.866$ & $1$ & $1$ & $-0.4$ & $-$ & $1$ & $1$ & $12$ & $1.55$ \\
     Dymnikova & $\gamma = 0.892$ & $0.700$ & $2.000$ & $1$ & $1$ & $0$ & $-$ & $1$ & $1$ & $12$ & $1.5$ \\
     AS-collapse & $\xi= 0.167 $ & $0.448$ & $1.858$ & $1$ & $1$ & $-0.09$ & $-1.18$ & $1$ & $1$ & $12$ & $1.5$ \\
\end{tabular}
\end{ruledtabular}
\caption{\label{tab:initflow} Settings used for generating the stream plots illustrating the phase space dynamics in Fig.\ \ref{Fig.phasespace}.}
\end{table*}
\noindent
\emph{Bardeen black holes:} \\
We start by discussing the Bardeen black hole \eqref{geo.Bardeen}. This leads to the simplest phase space structure. Substituting \eqref{geo.Bardeen} into eq.\ \eqref{eq.mp.1} gives
\be\label{eq.mp.Bardeen}
\dot{m}_+ = \frac{1}{R^3} \left[ (R^2 + a^2)^{3/2} - 2 m_+ R^2 \right] \, F(v) \, , 
\ee
with 
\be\label{eq.Fv.Bardeen}
F(v) = \frac{R^3}{\left(R^2 + a^2 \right)^{3/2} - 2 \, m_- \, R^2} \, \frac{(p-1) \, \beta}{v^p} \, .  
\ee 
The RHS of eq.\ \eqref{eq.mp.Bardeen} is linear in $m_+$. Thus there is one value for $m_{+}$ that leads to a vanishing RHS when $R(v)$ is given by $r_-$:
\be\label{eq.pos.qfp.Bardeen}
m_{+,*}^{\rm rep} = \frac{1}{2 r_-^2} \left( r_-^2 + a^2 \right)^{3/2} \, . 
\ee 
This quasi-fixed point governs the late-time behavior of the dynamics. The stream plot shown in the top-left diagram of Fig.\ \ref{Fig.phasespace} reveals that the quasi-fixed point is repulsive (rep). The mass function $m_+(v)$ diverges as $v \rightarrow \infty$. The endpoint is either 
\be 
\lim_{v \rightarrow \infty} m_+(v) = \pm \infty \, .
\ee

Notably, the RN geometry \eqref{geo.RN} leads to the same phase space structure. So we are not discussing this case separately. \\ 

\noindent
\emph{Hayward black holes:} \\
We proceed with the discussion of the Hayward geometry \eqref{geo.Hayward}. In this case, the evaluation of eq.\ \eqref{eq.mp.1} yields
\be\label{eq.mp.Hayward}
\dot{m}_+ = \frac{[R^3 - 2 (R^2-l^2) \, m_+ ] (R^3 + 2l^2 m_+)}{R^6} \, F(v) \, , 
\ee 
with
\be
F(v) = \frac{R^6}{(R^3 + 2l^2 \mminus) [R^3 - 2 \mminus (R^2-l^2)]} \, \frac{(p-1)\beta}{v^p} \,  .
\ee
In contrast to the Bardeen black hole, the RHS of eq.\ \eqref{eq.mp.Hayward} is quadratic in $m_+$. This implies that there are two quasi-fixed points where the RHS vanishes when evaluated at $R=r_-$:
\begin{subequations}
\begin{align}\label{eq.rp.Hayward}
m_{+,*}^{\rm rep} & = \frac{r_{-}^3}{2(r_{-}^2-l^2)}>0 \, , \\
\label{eq.at.Hayward}
m_{+,*}^{\rm att} & =  -\frac{r_{-}^3}{2l^2} < 0 \, .
\end{align}
\end{subequations}
The stream plot shown in the top-right diagram of Fig.\ \ref{Fig.phasespace} shows that the first point is repulsive (rep) while the second one acts as an attractor (att) as $v \rightarrow \infty$.

At this point, we notice the following feature of the phase space. Since the RHS of \eqref{eq.mp.Hayward} is quadratic in $m_+$, trajectories reach $m_+(v) = + \infty$ at finite values $v$. The Misner-Sharp mass \eqref{geo.Hayward} is finite and continuous when going from $m_+ = \infty$ to $m_+ = - \infty$. This indicates that the trajectories can ``go out at the top'' and ``return at the bottom'' of the $R$-$m_+$$-$plane. In other words the phase space of the Hayward model is not a plane but actually a cylinder. Thus all streams ultimately reach the late-time attractor \eqref{eq.at.Hayward}. The attractor is global, and its induced scaling properties hold independent of the initial conditions. \\

\noindent
\emph{Dymnikova black holes} \\
The evaluation of \eqref{eq.mp.1} for the mass function of the Dymnnikova black hole \eqref{geo.Dymnikova} leads to
\be\label{eq.mp.Dymnikova}
\dot{m}_{+}(v) = \frac{\gamma^2 \mplus}{R} \, \frac{\left[ R - 2 \mplus \left( 1-e^{-\frac{R^{3}}{\gamma^2 \mplus}} \right) \right]}{\left[ \gamma^2 m_+ -  e^{-\frac{R^3}{\gamma^2 m_+}}  \left( R^3 + \gamma^2 m_+ \right) \right]} F(v)
\ee
with
\be
\begin{split}
F(v) = & \frac{R}{\gamma^2 m_-} \, \frac{1}{R-2m_- \left( 1 - e^{-\frac{R^3}{\gamma^2 m_-}} \right)} \\ & \times \left[ \gamma^2 m_- -  e^{-\frac{R^3}{\gamma^2 m_-}}  \left( R^3 + \gamma^2 m_- \right) \right] \, \frac{(p-1)\beta}{v^p} \,  .
\end{split}
\ee
Again, the mass function $M(r)$ is continuous and finite at $m \rightarrow \pm \infty$, so that solutions can leave the phase space on the top and reenter at the bottom. Thus we again encounter the cylinder structure realized in the Hayward model. The Dymnikova geometry gives rise to a qualitatively new feature though. In this case the mass function \eqref{geo.Dymnikova} possesses a discontinuity at $m=0$:
\be\label{Dym.discont1}
\begin{split}
& \lim_{m \rightarrow 0^+} M(r) = 0 \, , \\
& \lim_{m \rightarrow 0^-} M(r) = + \infty \, . 
\end{split}
\ee
This discontinuity propagates into the RHS of eq.\ \eqref{eq.mp.Dymnikova} where
\begin{align}\label{Dym.discont1m}
& \lim_{m_+ \rightarrow 0^+} \dot{m}_{+}(v) = F(v) \, , \\
& \lim_{m_+ \rightarrow 0^-} \dot{m}_{+}(v) = 0 \, . 
\end{align}
This has the profound consequence that solutions cannot cross the line $m_+(v) = 0$ and terminate at this discontinuity. In terms of the phase space this entails that there is a new class of solutions which terminate at finite $v$. These solutions correspond to the brown cluster in the phase space diagram.

We proceed by finding the quasi-fixed points of eq.\ \eqref{eq.mp.Dymnikova}. Again there are two roots where the RHS vanishes. The first root arises as the solution of the transcendental equation,
\be\label{eq.rp.Dymnikova} 
r_- - 2 m_{+,*}^{\rm rep} \bigg( 1-e^{-\frac{r_-^{3}}{\gamma^2 m_{+,*}^{\rm rep}}} \bigg)=0 \, . 
\ee
Since $r_- > 0$ this root is located at $m_{+,*}^{\rm rep} > 0$. The repulsive property of this quasi-fixed point is readily deduced from the phase diagram. In addition, there is the attractor
\be\label{eq.ap.Dymnikova}
m_{+,*}^{\rm att}  =  0 \, . 
\ee 
Owed to the discontinuity \eqref{Dym.discont1m} the attractor is one-sided only: it acts as an attractor for solutions approaching zero from below only. This feature discriminates the attractor from the global attractor found in the Hayward case. \\[1ex]

\noindent
\emph{Black holes from the AS-collapse} \\
Eq.\ \eqref{eq.mp.1} evaluated for the AS-collapse \eqref{geo.log} is
\begin{equation}
\label{eq.mp.collapse}
\begin{split}
& \dot{m}_{+}(v)  = \\
& \; \left[1-\frac{R^2}{6\xi}\log \bigg( 1+\frac{6\xi \mplus}{R^3} \bigg)^2 \right] \left( 1+\frac{6\xi \mplus}{R^3} \right) \, F(v) \, , 
\end{split}
\end{equation}
with
\begin{equation}
\label{eq.Fv.collapse} 
F(v) = \frac{1}{\left[ 1-\frac{R^2}{6\xi}\log \big( 1+\frac{6\xi \mminus}{R^3} \big)^2 \right] \left(1+\frac{6\xi \mminus}{R^3} \right)} \, \frac{(p-1) \beta}{v^p} \, . 
\end{equation}
The search for quasi-fixed points reveals that there are three values where the RHS of \eqref{eq.mp.collapse} vanishes when evaluated at $R=r_-$:
\begin{subequations}\label{eq.QFPs.collapse}
\begin{align}\label{mp.rep1.AScollapse}
m_{+,*}^{\rm rep} = & - \frac{r_{-}^3}{6 \xi} \bigg(1 -  e^{\frac{3\xi}{r_{-}^2}}  \bigg) >0 \, , \\ \label{mp.att1.AScollapse}
m_{+,*}^{\rm att} = & -\frac{r_{-}^3}{6 \xi} < 0 \, , \\  \label{mp.rep2.AScollapse}
m_{+,*}^{\rm rep} = & - \frac{r_{-}^3}{6 \xi} \bigg(1 +  e^{\frac{3\xi}{r_{-}^2}} \bigg)  < 0 \, .  
\end{align}
\end{subequations}
The attractive and repulsive properties of these points are readily inferred from the stream plot shown in the bottom-left diagram of Fig.\ \ref{Fig.phasespace}, justifying the superscripts. Eq.\ \eqref{eq.QFPs.collapse} moreover shows that the repulsive quasi-fixed points sandwich the attractor. As a consequence, the streams dragged into the attractor are located in a wedge, while generically one finds that $m_+$ diverges as the shell settles on the CH. In contrast to the Hayward case, the streams cannot be connected at infinity, so that the phase space of the AS-collapse model is a plane. \\
%

\subsection{quasi-Fixed Points: Analytic Results}
\label{Sect.3.1}
In the previous section, we identified several late-time attractors for the various regular black hole geometries studied in this work. The goal of this section is to obtain analytic results for the scaling behavior associated with these attractors. Throughout this section, we will use the $\simeq$ symbol to stress that an expression corresponds to an asymptotic expansion at large values of $v$ and it is tacitly understood that subleading corrections to the expansion are not displayed. The discussion follows the same order as the previous section. For convenience, our findings are summarized in Table \ref{Tab.summary_relevant}. 
\begin{table*}[htbp!]
\begin{ruledtabular}
\renewcommand{\arraystretch}{1.7}
\begin{tabular}{l|l|ll|lll}
\textit{Solution} & \textit{Topology} & \multicolumn{2}{c|}{$M_{+}(v)$} & $K_+(v)$ & $C^2_+(v)$ & $\Psi_{2+}(v)$  \\ \hline \hline
     Bardeen  & plane  &  & $\propto \pm v^{-p}e^{\kappa_{-}v}$ & $\propto v^{-2p} \, e^{2 \kappa_- v}$ & $\propto v^{-2p} \, e^{2 \kappa_- v}$ & $\propto \pm v^{-p} \, e^{\kappa_- v}$ \\ \hline
     Hayward  & cylinder & & $\propto v^{p}$ & $\propto v^{6p}$ & $\propto v^{6p}$ & $\propto -v^{3p}$ \\ \hline
     \multirow{2}{*}{Dymnikova}& \multirow{2}{*}{cylinder} & Case I: & $\propto v^{-(p+1)}e^{\kappa_{-}v}$ 
     & $\propto v^{-2(p-1)} \, e^{2\kappa_- v}$ & $\propto v^{-2(p-1)} \, e^{2\kappa_- v}$ & $\propto - v^{-(p-1)} \, e^{\kappa_- v}$
     \\ 
     & & Case II: & $M_{+}(v_{*}) \to 0^{+}$ & - & - & -   \\ \hline
  \multirow{2}{*}{AS-collapse}& \multirow{2}{*}{plane}    &  Case I: & $\propto  v^{-p}e^{\kappa_{-}v}$ & $\propto v^{-2p} \, e^{2 \kappa_- v}$ & $\propto v^0$ & $\propto  v^0$   \\ 
  & & Case II: & $\propto-\log v^{p}$ & $\propto v^{4p}$ & $\propto v^{4p}$ & $\propto v^{2p}$  \\
\end{tabular}
\end{ruledtabular}
\caption{\label{Tab.summary_relevant} Summary of the key results found in Sect.\ \ref{Sect.3}. The column ``Topology'' indicates whether solutions reach $m_+ = \infty$ at finite $v$ and can be continued at $m_+ = -\infty$ (cylinder) or end at finity (plane). The asymptotic scaling of the Misner-Sharp mass is summarized in the column ``$M_+(v)$'' while the columns ``$K_+(v)$'', $C_+^2(v)$'', and ``$\Psi_{2+}(v)$'' give the asymptotics of the Kretschmann scalar, the square of the Weyl-tensor and the Coulomb-component of the Weyl-tensor. All singularities are weak in the Tipler sense. Polynomial growths of the curvature invariants furthermore implies that they are also weak in the Królak sense.}
\end{table*}

We start with the late-time dynamics of the shell impacting on the Cauchy horizon. Adapting the general analysis of Sect.\ \ref{Sect.2.2} to the geometries \eqref{eq:geometries}, one readily finds that the late-time dynamics of $R(v)$ is model-independent to a large extend. The terms relevant to the construction of the Misner-Sharp mass read
\be\label{RHay}
R(v) - r_- \simeq  \, \frac{1}{v^{p-1}} \, a_0 \,   
\left[ 1 + \frac{p-1}{\kappa_- v } + \frac{p(p-1)}{\kappa_-^2 v^2} \right] \, . 
\ee
The constants $a_0$ capture the information on the geometry. Their explicit values are readily obtained from evaluating eq.\ \eqref{a0coeff} and read
\begin{subequations}
\begin{align}
\text{Bardeen:} \; & a_0 =\frac{\beta}{\kappa_{-}} \frac{\, r_-^2}{\left(a^2+r_-^2\right)^{3/2}} \, , \\
\text{Hayward:} \; & a_0 =\frac{\beta}{\kappa_{-}} \frac{r_-}{4 \, m_0^2} \, , \\
\text{Dymnikova:} \; & a_0 =\frac{\beta}{\kappa_{-}} \frac{\left(r_-^3 + m_0 \, \gamma^2-2 m_0 \, r_-^2\right)}{2 \, \gamma^2 \ m_0^2} \, , \\
\text{AS-collapse:} \; & a_0 =\frac{\beta}{\kappa_{-}} \frac{1}{ r_-} \, \left(1 + \frac{6 m_0 \xi }{r_-^3}\right)^{-1} \, . 
\end{align}
\end{subequations}

Analyzing the dynamics related to $m_+(v)$ then requires the late-time asymptotics of $F(v)$. Retaining terms up to $a_2$ in the asymptotic expansion of $R(v)$, one finds that
\be\label{eq.Bar.Fasym}
F(v) \simeq - \frac{r_-}{2} \left( \kappa_- - \frac{p}{v} \right) \, . 
\ee
The relative factor between the terms appearing in the brackets are set by the recursion relation \eqref{ancoeff}. Remarkably, this result holds for all geometries in this section. The resulting consequences for the Misner-Sharp mass depend on the chosen geometry, though, and are discussed on a case by case basis.\\[1ex]

\noindent
\emph{Bardeen black holes} \\
We start by investigating the consequences of the quasi-fixed point \eqref{eq.pos.qfp.Bardeen} leading to the phase diagram shown in the top-right corner of Fig.\ \ref{Fig.phasespace}. The repulsive nature of the quasi-fixed point indicates that, asymptotically, $m_+$ will be large. Thus, we can approximate \eqref{eq.mp.Bardeen} by retaining the terms proportional to $m_+$ on its RHS only. Substituting the asymptotic expansion \eqref{eq.Bar.Fasym} then yields the following approximation, capturing the late-time behavior of $m_+(v)$:
\be\label{bardeen.mp.late}
\dot{m}_+ \simeq m_+ \left( \kappa_- - \frac{p}{v} \right) \, . 
\ee
This equation is readily integrated, yielding
\be\label{mp.Bar.asym}
m_+(v) \simeq \pm c \, v^{-p} \, e^{\kappa_- v} \, .
\ee
Here $c > 0$ is an integration constant. Note that the result applies universally for flows being repelled by $m_{+,*}^{\rm rep}$ with the signs in \eqref{mp.Bar.asym} capturing the limits $m_+(v) = \pm \infty$.  Eq.\ \eqref{mp.Bar.asym} allows to reconstruct the asymptotics of the Misner-Sharp mass, establishing that 
\be\label{mass-inflation-instability}
M_+(v) \simeq \, \pm \frac{r_-^3}{(r_-^2 + a^2)^{3/2}} \, c \, v^{-p} \, e^{\kappa_- v} \, . 
\ee
Hence $M_+(v)$ grows exponentially and we establish that the repulsive quasi-fixed point yields the standard result of eternal mass-inflation.  Again, the result \eqref{mp.Bar.asym} also holds for the RN geometry. \\[1ex]

\noindent
\emph{Hayward black holes} \\
The Hayward geometry is a prototypical realization of the generic scaling behavior discussed in Sect.\ \ref{Sect.2.2}. Its late-time dynamics is governed by the quasi-fixed point \eqref{eq.at.Hayward} which acts as a  global attractor. Thus all solutions ultimately enter the corresponding scaling regime, independently of their initial conditions. 

The scaling behavior of $m_+(v)$ is readily obtained from the Frobenius analysis. Including all terms relevant for the reconstruction of $M_+(v)$ one has
\be\label{mpHay}
m_+(v) - m_{+,*}^{\rm att} \simeq - \frac{3 \, \beta \, r_-^3}{8 \, l^2 \, m_0^2 \, \kappa_-} \frac{1}{v^{p-1}} \left[ 1 + \frac{2 (p-1)}{\kappa_- v} \right] \, . 
\ee
Note that there is no free coefficient in the expansion. Thus the attractor fixes the late-time behavior completely. Substituting the asymptotic expansions \eqref{RHay} and \eqref{mpHay} into the Misner-Sharp mass \eqref{geo.Hayward} and taking into account the cancellation of the leading terms appearing in the denominator finally yields 
\be \label{MisnerAttractorHY}
M_{+}(v) \simeq \frac{2}{3} \frac{r_{-}^3 \, m_{0}^2 \, \kappa_{-}^2}{\beta \, (p-1) \, l^2}  \,  v^{p} \, .
\ee
Notably, $M_+(v)$ no longer grows exponentially in $v$ and follows a power-law behavior. This is the imprint of the attractor discovered in \cite{Bonanno:2020fgp,Bonanno:2023qhp}. In combination with the phase-space analysis, this entails that any initial condition will ultimately follow this power law. \\[1ex]

\noindent
\emph{Dymnikova black holes} \\
Case I: The late-time attractor shown in the bottom-left diagram of Fig.\ \ref{Fig.phasespace} is characterized by the limit $m_+ \rightarrow 0^-$. In this limit, the right-hand side of eq.\ \eqref{eq.mp.Dymnikova} is dominated by the exponential terms. Retaining these contributions only, the exponentials in the numerator and denominator cancel. Substituting the late-time behavior of $F(v)$ given by \eqref{eq.Bar.Fasym} then leads to the following differential equation capturing the late-time behavior of $m_+(v)$ close to the one-sided attractor
\be\label{Dym.att.dyn}
\dot{m}_+ \simeq \frac{\gamma^2 \, \kappa_-}{r_-^3} \left( 1 - \frac{p}{\kappa_- v} \right) \, m_+^2 \, . 
\ee
This equation is readily integrated, yielding
\be\label{Dyn.att.sol}
m_+(v) \simeq - \frac{r_-^3}{\gamma^2} \left( \kappa_- \, v - p \log v + \bar{c}\right)^{-1} \, , 
\ee
where $c$ is an integration constant. Substituting this result into the Misner-Sharp mass \eqref{geo.Dymnikova} then yields
\be\label{Dyn.ms.asym}
M_+(v) \simeq c \, \frac{r_-^3}{\gamma^2 \, \kappa_-} \, v^{-(p+1)} \, e^{\kappa_- v} \, ,
\ee
with $c>0$ always. Thus, the late-time attractor \eqref{eq.ap.Dymnikova} entails mass-inflation instability. The shift in the exponent appearing in the subleading power-law corrections thereby arises from the multiplicative factor $m$ appearing in the Misner-Sharp mass. \\[1ex]

\noindent
Case II: The case of trajectories terminating at $m_+(v_*) = 0^+$ will not be discussed in detail and is added to Table \ref{Tab.summary_relevant} for completeness. \\[1ex]

\noindent
\emph{Black holes from the AS-collapse} \\
The analysis of the late-time behavior of the AS-collapse model proceeds along similar lines. Based on the structure of the phase space, we distinguish two cases: first, there are the repulsive flows associated with the quasi-fixed points \eqref{mp.rep1.AScollapse} and \eqref{mp.rep2.AScollapse}. These lead to solutions where $m_+(v) \rightarrow \pm \infty$, respectively. Second, one has solutions whose late-time behavior is controlled by the attractive quasi-fixed point \eqref{mp.att1.AScollapse}. We discuss these cases in turn. \\[1ex]

\noindent
Case I: mass-inflation instability from $m_{+,*}^{\rm rep}$: \\
For the repulsive quasi-fixed points we start from \eqref{eq.mp.collapse} and consider the regime where $m_+ \gg 1$. Retaining the leading and subleading terms contributed by $F(v)$, the asymptotic behavior of these solutions at large values $m_+$ is captured by
\be\label{AS-asmp-case1}
\dot{m}_+ \simeq \frac{\kappa_-}{2} \log(m_+^2) \, m_+ \left( 1 - \frac{p}{\kappa_- v} \right) \, . 
\ee
This approximation is at the same level as eq.\ \eqref{bardeen.mp.late}, capturing the mass-inflation instability in the Bardeen geometry. Notably, eq.\ \eqref{AS-asmp-case1} admits an analytic solution
\be\label{AS-rep-sol}
m_+(v) \simeq \exp\left(c \, v^{-p} e^{\kappa_- v } \right) \, , 
\ee
where $c$ is an integration constant. Reconstructing the Misner-Sharp mass based on \eqref{AS-rep-sol} yields
\be\label{AS-Misner-Sharp-sol-rep}
M_+^{\rm rep} \simeq \frac{r^3_-}{6 \xi} \, |c| \, v^{-p} \, e^{\kappa_- v} \, . 
\ee
 Remarkably, the repulsive quasi-fixed points lead to the standard mass-inflation instability. Since the Misner-Sharp mass contains the square of $m_+(v)$ only, the result \eqref{AS-Misner-Sharp-sol-rep} applies to both cases $m_+(v) = \pm \infty$. \\[1ex]

\noindent
Case II: logarithmic growth of $M_+(v)$ induced by $m_{+,*}^{\rm att}$: \\
The derivation of the late-time behavior induced by the attractive quasi-fixed point is slightly more complicated and its details are relegated to Appendix \ref{App.D}. In summary, the attractor imprints a logarithmic growth of the Misner-Sharp mass
\be\label{result.main}
M_+^{\rm att}(v) \simeq \frac{r_-^3}{12 \xi} \log\left( y_0 \, v^{-p} \right)^2 \, .  
\ee
The constant $y_0$ is the solution of the transcendental equation
\be\label{y0-equation} 
3a + b\ y_0 + \frac{1}{2} \kappa_- \, y_0 \log(y_0)^2 = 0 \, . 
\ee
Here the constants $a$ and $b$ encode properties of the background geometry and are defined in eq.\ \eqref{ab.def}. 

It is remarkably that the AS-collapse model contains a phase where the build-up of the Misner-Sharp mass at the inner horizon is even slower than power law. This discovery constitutes the main finding of our work.

\subsection{Numerical Analysis}
\label{Sect.3.2}
We continue the investigation of the Ori-model by presenting explicit examples obtained by solving the coupled system of differential equations \eqref{eq.Rdiv} and \eqref{eq.mp.1} numerically. The results corroborate the analytical results derived in the last subsection. Moreover, they show that the late-time attractors are indeed reached also from initial conditions placed outside of the scaling regimes governed by the quasi-fixed points. The typical time-scale for reaching the attractor is set by the surface gravity $\kappa_-$ which is set to unity in all our examples. Throughout this section, we restrict the discussion to the mass $m_+(v)$ and the Misner-Sharp mass $M_+(v) \equiv M_+[R(v), m_+(v)]$. The later is constructed by evaluating \eqref{eq:geometries} on the numerical solutions for $\{R(v), m_+(v)\}$.\footnote{The construction of $M_+(v)$ requires solving the coupled system of differential equations with high numerical precision. This applies in particular to the AS-collapse model where one encounters cancellations among leading contributions within the argument of the logarithm.} We do not plot results for the curvature scalars \eqref{Ricci-scalar}-\eqref{Weyl}. Their  scaling behavior readily follows from the one of $M_+(v)$. Since there are no cancellations between the single terms, this is straightforward and does not lead to additional insights.

\begin{figure*}[!p]
\includegraphics[width = 0.42\textwidth]{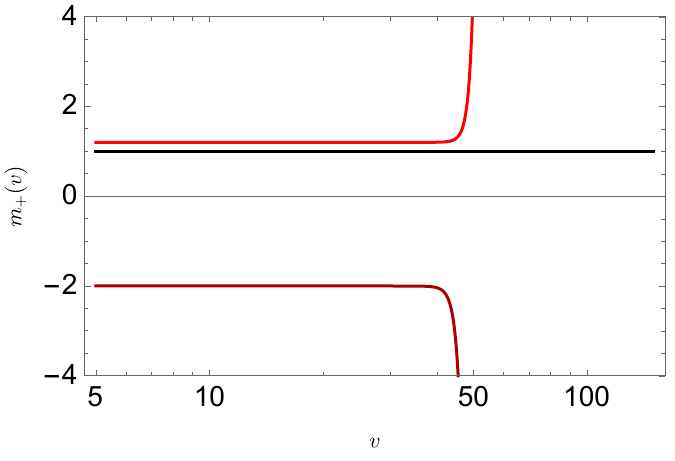} \, \,
\includegraphics[width = 0.45\textwidth]{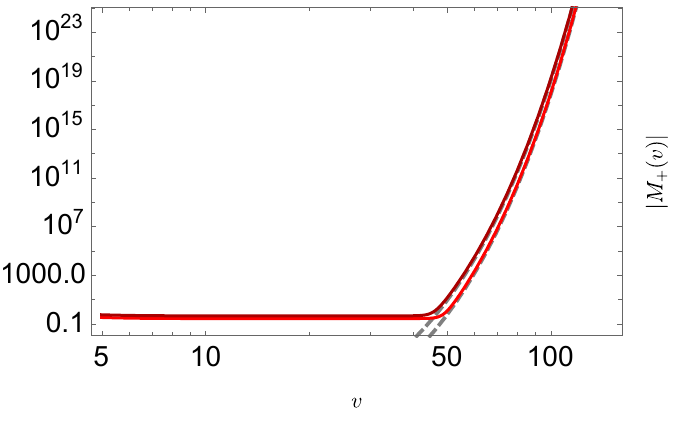} \\
\includegraphics[width = 0.42\textwidth]{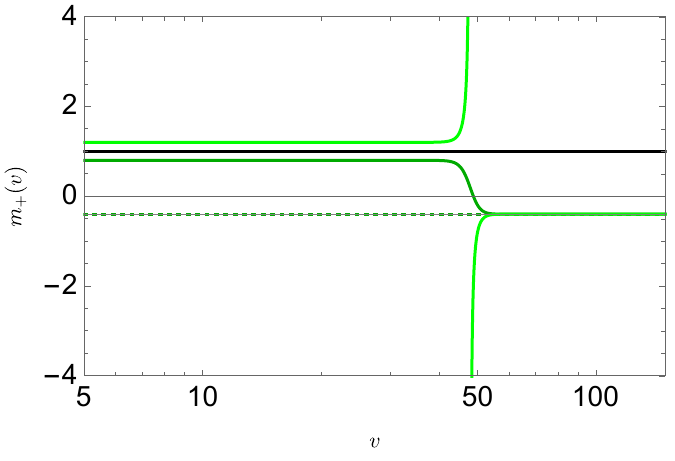} \, \, 
\includegraphics[width = 0.45\textwidth]{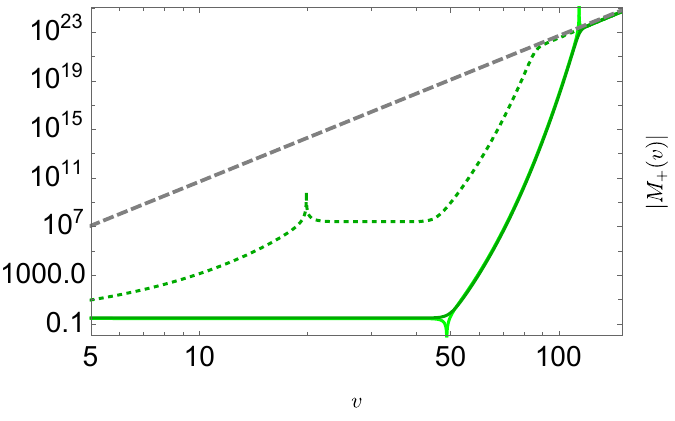} \\
\includegraphics[width = 0.42\textwidth]{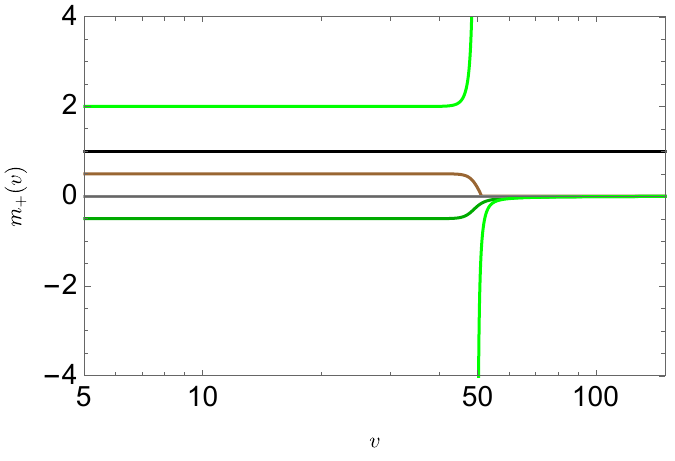} \, \,
\includegraphics[width = 0.45\textwidth]{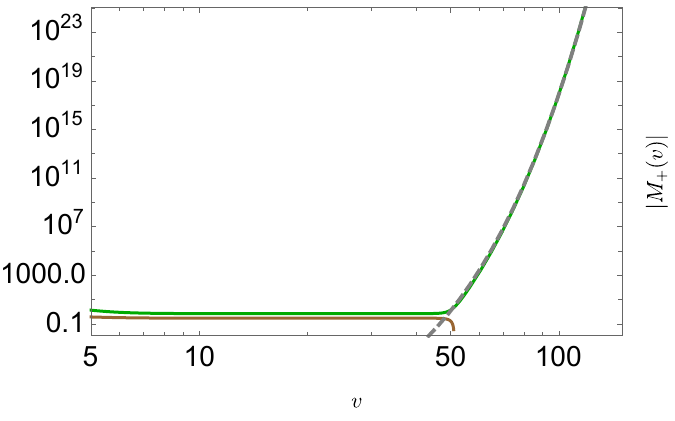} \\
\includegraphics[width = 0.42\textwidth]{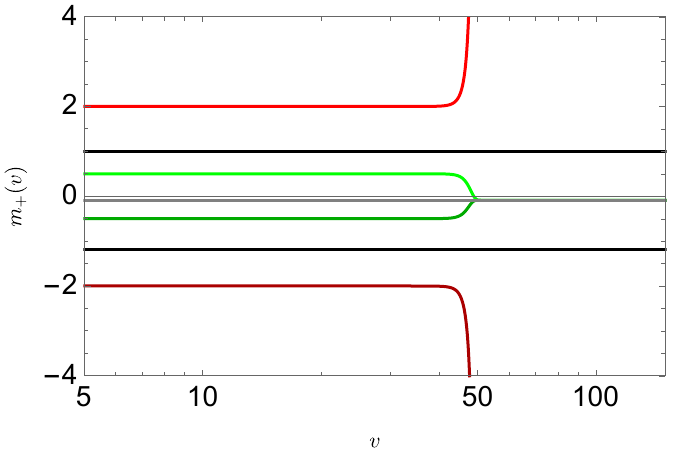} \, \,
\includegraphics[width = 0.45\textwidth]{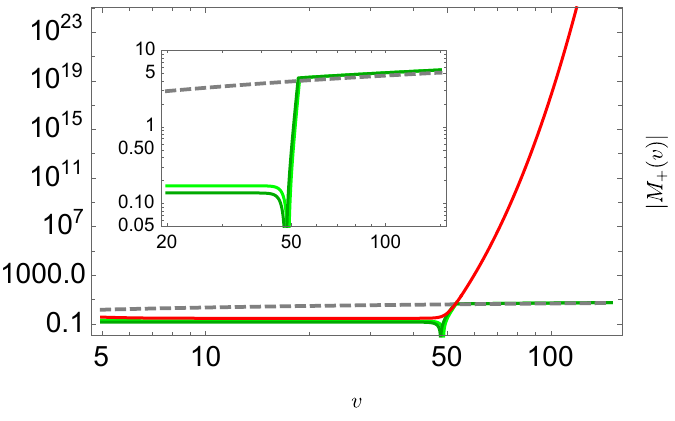} \\
\caption{\label{Fig.examples} Illustration of the dynamics found for the Ori-model for various regular black hole geometries. From top to botton the geometry is given by the Bardeen black hole \eqref{geo.Bardeen}, the Hayward geometry \eqref{geo.Hayward}, the Dymnikova black hole \eqref{geo.Dymnikova}, and the AS-collapse \eqref{geo.log}, respectively. The left column shows the mass-function $m_+(v)$ while the right column displays the absolute value of the Misner-Sharp mass $M_+(v)$ reconstructed from the numerical solutions for $\{R(v), m_+(v)\}$ on a double-logarithmic scale. The attractive (repulsive) quasi-fixed points for $m_{+,*}$ are given by gray (black) horizontal lines while the late-time attractors for $M_+(v)$ constructed in Sect.\ \ref{Sect.3.1} are superimposed as gray dashed lines. See the main text for further details.}
\end{figure*}
%

%
\begin{table*}[!t]
\begin{ruledtabular}
\renewcommand{\arraystretch}{1.5}
\begin{tabular}{l|l|cc|c}
     \textit{Solution} & \textit{Final $M_{+}(v)$} & $R(v_{\text{i}})$ & $m_+(v_{\text{i}})$ & $c$ \\ \hline \hline
     \multirow{2}{*}{Bardeen} & mass-inflation & $0.5$ & $\phantom{-} 1.2$ & $\phantom{-}0.146$ \\
     & mass-inflation & $0.5$ & $-2$ & $-2.197$\\ \hline
     \multirow{3}{*}{Hayward} & polynomial attractor & $0.58$ & $0.8$ & $-$ \\
     & polynomial attractor & $0.58$ & $m_{+,*}^{\rm att}$ & $-$ \\
     & polynomial attractor & $0.58$ & $1.2$ & $-$ \\ \hline
     \multirow{3}{*}{Dymnikova} & mass-inflation & $0.8$ & $\phantom{-}2$ & $0.102$\\ 
     & terminates at $v_*$ & $0.8$ & $\phantom{-}0.5$ & $-$ \\ 
     & mass-inflation &  $0.8$ & $-0.5$ & $1.133$ \\ \hline
     \multirow{4}{*}{AS-collapse} & mass-inflation & $0.5$ & $\phantom{-}2$ & $8.99 \times 10^{-3}$\\
     & log attractor & $0.5$ & $\phantom{-}0.5$ & $-$ \\
     & log attractor & $0.5$ & $-0.5$ &  $-$\\
     & mass-inflation & $0.5$ & $-2$ &  $0.01$ \\
\end{tabular}
\caption{\label{tab:inittrajectories} Initial conditions used for generating the numerical integrations in Fig.\ \ref{Fig.examples}. All values are imposed at $v_{\text{i}}=5$. The background geometries are fixed by settings in Tab.\ \ref{tab:initflow}. If appropriate, the value of the free integration constant $c$ appearing in relation to the repulsive quasi-fixed points is added in the column $c$.}
\end{ruledtabular}
\end{table*}
We proceed by discussing the examples shown in Fig.\ \ref{Fig.examples}, which are obtained from  solving the Ori-model on different regular black hole backgrounds numerically. The color code used in all examples is as follows: solutions attracted (repelled) by a quasi-fixed point are coded in green (red), and we reserve brown for solutions of the Dymnikova model which terminate at a finite value $v_*$. Throughout the figure, the values for $m_+$ corresponding to the attractive (repulsive) quasi-fixed points are included as gray (black) horizontal lines. Relevant attractors are superimposed as gray dashed lines. The parameters specifying the regular black hole geometries are again the ones given in Table \ref{Fig.phasespace}. The initial conditions generating the specific examples are collected in Table \ref{tab:initflow}. \\

\noindent
\emph{Bardeen geometry} (first line of Fig.\ \ref{Fig.examples}): \\
The dynamics of $m_+(v)$ arising from the Bardeen geometry is displayed in the left diagram. The repulsive quasi-fixed point \eqref{eq.pos.qfp.Bardeen} has been added as the black horizontal line. The solutions are driven away from this line and their dynamics follows the asymptotics \eqref{mp.Bar.asym}. The explicit value of the integration constant $c$ is obtained by equating the numerical solution to the asymptotics at a fixed value $v$ and list in column ``$c$'' of Table \ref{tab:inittrajectories}. The absolute value of the Misner-Sharp mass evaluated at the position of the shell is shown in the right diagram. Again the asymptotic late-time behavior follows the relation \eqref{mass-inflation-instability} which is superimposed as the gray-dashed lines. This establishes that both solutions exhibit eternal mass-inflation. \\

\noindent
\emph{Hayward geometry} (second line of Fig.\ \ref{Fig.examples}): \\
The dynamics resulting from the Hayward geometry is shown in the second line of Fig.\ \ref{Fig.examples}. The solutions $m_+(v)$ shown in the left panel are governed by the interplay of the repulsive quasi-fixed point \eqref{eq.rp.Hayward} and the global attractor \eqref{eq.at.Hayward}, highlighted, respectively, as the black and gray horizontal line. Solutions for $m_+(v)$ starting above the repulsive line reach infinity at a finite value of $v$ and subsequently reenter at minus infinity before reaching the global attractor from below. Solutions starting between the two lines approach the attractor from above. The reconstruction of $M_+(v)$ at the position of the shell is shown in the right diagram. Both solutions approaching $m_{+,*}^{\rm att}$ from above and below undergo a phase of transient mass inflation before entering the polynomial scaling regime induced by the attractive quasi-fixed point. Notably, the attractor behavior for $M_+(v)$ is universal in the sense that it does not contain any free parameter. In other words, the late-time behavior is independent of the initial condition or the history taken by $M_+(v)$ before entering the attractor regime. In particular, one can find special solutions which start tracking the polynomial attractor early-on, so that the episode of transient mass-inflation is short. An example for this is the solution added as the dashed green line. The trajectory starts on $m_{+,*}^{\rm att}$ and leads to the curve $M_+(v)$ reaching the attractor first. \\

\noindent
\emph{Dymnikova geometry} (third line of Fig.\ \ref{Fig.examples}): \\
Similarly to the Hayward geometry, the dynamics of $m_+(v)$ in the Dymnikova case is dictated by the interplay of the repulsive quasi-fixed point \eqref{eq.rp.Dymnikova} and attractive quasi-fixed point \eqref{eq.ap.Dymnikova}. Solutions above the repulsive line leave the phase space at the top and reenter at the bottom before reaching the attractor from below. Solutions starting between the lines approach the attractor from above and reach $m_{+,*}^{\rm att} = 0$ at finite $v_*$. This new feature is highlighted by the brown line. The Misner-Sharp mass evaluated along the solutions $\{R(v), m_+(v)\}$ is shown in the right panel. The numerical examples establish that solutions reaching $m_{+,*}^{\rm att}$ from below follow the asymptotic behavior \eqref{Dyn.ms.asym} with the integration constants $c$ given in Table \ref{tab:inittrajectories}. Thus we conclude that all solutions of the Dymnikova geometry reaching $v= \infty$ exhibit eternal mass inflation. \\

\noindent
\emph{AS-collapse geometry} (fourth line of Fig.\ \ref{Fig.examples}): \\
Finally, the AS-collapse model is detailed in the bottom row of Fig.\ \ref{Fig.examples}. The three quasi-fixed points \eqref{eq.QFPs.collapse} governing the dynamics of $m_+(v)$ are again shown as gray and black horizontal lines. Solutions above and below the repulsive lines go off to infinity. In addition, there are the solutions sandwiched between the repulsive lines which asymptote to the attractor $m_{+,*}^{\rm att}$. Reconstructing $M_+(v)$, we find that the first class of solutions follow the asymptotics \eqref{AS-Misner-Sharp-sol-rep} and lead to eternal mass inflation. The second class leads to a logarithmic growth of $M_+(v)$ following \eqref{result.main}. The constant $y_0$ determining the position of the attractor is obtained by solving eq.\ \eqref{y0-equation} numerically, yielding $y_0 = 31.549$. In this context, it is worth highlighting two remarkable features. First, the attractor does not come with a free parameter. Hence the asymptotics is universal and independent of the initial conditions as long as the solution is within this class. Second, the attractor can be reached without an appreciable episode of transient mass inflation. Thus the log-attractor tames the mass-inflation instability completely. This constitutes the key discovery of our work. \\

\begin{figure}[t]
\begin{center}
\includegraphics[width = 0.45\textwidth]{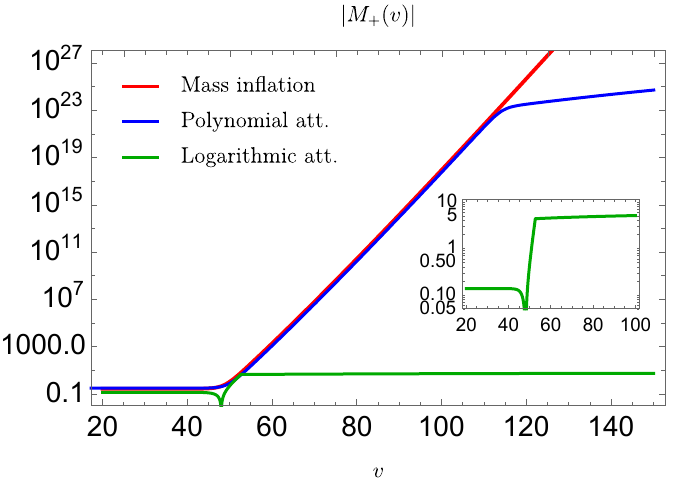}
\end{center}
\caption{\label{Fig.examples2}  Comparison, in a log-plot, between different dynamics of the Ori-model found within the phase space of regular black hole geometries. The examples cover eternal mass inflation (red line), the polynomial attractor realized by the Hayward model (blue line) and the log-attractor \eqref{mp.att1.AScollapse} of the AS-collapse model (green line). While solutions still undergo transient mass inflation before reaching the polynomial attractor, this feature is absent in the log-attractor case.}
\end{figure}
We close this section with the following observation. A remarkable outcome of our phase-space analysis is that, despite substantial differences in the geometry, there are just three different types of asymptotic scaling behavior of the Misner-Sharp mass. These are summarized in Fig.\ \ref{Fig.examples2}.  They comprise eternal mass inflation (red) where $M_+(v)$ grows exponentially in $v$, the polynomial attractor (blue) where $M_+(v)$ grows polynomial in $v$, and the log-attractor (green) where $M_+(v)$ grows logarithmically in $v$. 

\subsection{Singularity Strength}
\label{sect.6}
The asymptotic late-time behaviors determined in Sect.\ \ref{Sect.3.1} readily allow to determine the build-up of curvature singularities in the future sector of the shell. For this purpose, we evaluate the Kretschmann scalar $K$, the square of the Weyl-tensor $C^2$, and the Coulomb-component of the Weyl tensor $\Psi_2$ introduced in App.\ \ref{App.A}. The last two also serve as a measure for the tidal forces experienced by an observer. 

The results obtained from evaluating the general formulas \eqref{def:K}, \eqref{Weyl}, and \eqref{def:psi2} are listed in the last three columns of Table \ref{Tab.summary_relevant}. These reveals the following systematics. The exponential growth of $M_+(v)$ induces an exponential growth of the curvature scalars and $\Psi_2$. Analogously, the power-law scaling induced by the attractive quasi-fixed points in the Hayward and AS-collapse model also carries over to the curvature. The case I found in the AS-collapse is special since the Weyl-tensor based curvature invariance asymptote to constant, non-zero values. Comparing the Kretschmann scalar to the Weyl-curvature squared, one observes that the log-terms responsible for the exponential growth of the Kretschmann scalar are absent in the latter case.

At this stage it is instructive to relate these results ot the classification of singularities provided by Tipler \cite{Tipler:1977zza} and Królak \cite{Krolak:1986pno,Krolak:1987}. Following the analysis made in \cite{Bonanno:2020fgp}, one finds that one can always find a coordinate system which is regular in the future sector of the shell. As a consequence, all curvature singularities building up dynamically are weak in the Tipler sense.

Determining the strength of the singularity within the Królak classification requires the relation between the coordinate $v$ and the proper time $\tau$ measured by a massive observer moving along a radial geodesic in the future sector of the shell. It thereby turns out that, at asymptotically late times, this relation is given by
\be\label{vvstau}
\tau \propto e^{-\kappa_-v} \, ,
\ee
independently of the underlying dynamics of the model. This result has two profound consequences. First, the observer can reach $v=\infty$ in finite proper time. Hence the strength of the singularity is relevant when investigating whether a geodesic can be continued beyond this point. Second, the strength of the singularity can be determined by rewriting the scaling behavior in $v$ in terms of $\tau$. This establishes that the singularities related to an exponential growth of the curvature scalars is strong in the Królak definition. Conversely, power-law growth in $v$ indicates a weak singularity in this classification. We illustrate this based on the attractive quasi-fixed point appearing in the AS-collapse model. Rewriting the square of the Weyl-tensor in terms of $\tau$ yields
\be\label{sign-strength}
C^2_+ \propto (\log \tau)^{4p} \, . 
\ee
Thus $C^2_+$ becomes regular at $\tau = 0$ when it is integrated with respect to proper time once. This signals that the singularity is Królak weak and there may be a $C^1$-continuation of a geodesic beyond this point.
\section{Summary and Outlook}
\label{Sect.4}
In this work, we investigated the mass-inflation instability for several, representative regular black hole geometries including the geometries proposed by Bardeen, Hayward, Dymnikova, and the AS-collapse. Our phase space analysis of the Ori-model revealed that the stability of the geometry is governed by quasi-fixed points of the dynamical equations. These lead to a remarkable uniformity in the late-time dynamics. The Misner-Sharp mass determining the geometry in the vicinity of the Cauchy horizon exhibits just three distinct cases: exponential growth, polynomial growth, and logarithmic growth. The first case is characteristic for the mass-inflation instability while the second behavior originates from the attractors discovered in \cite{Bonanno:2020fgp}. The third case is uniquely realized by the regular geometries resulting from the gravitational collapse within the gravitational asymptotic safety program \cite{Bonanno:2023rzk}. The key discovery of our work is that in this case the log-attractor is reached without an appreciable phase of transient mass inflation. This distinguishes the AS-collapse geometry from other regular black hole geometries exhibiting polynomial attractors.

At the level of scalar curvature invariants, the three distinguished late-time behaviors for the Misner-Sharp mass reduce to two distinct scenarios. First the curvature may grow exponentially in $v$, signaling a mass-inflation instability. The resulting singularity is strong in terms of the Królak classification. Conversely, the polynomial- and log-attractors result in singularities which are weak in the Królak sense and may allow to extend geodesics beyond this point.

At this stage, we feel that our understanding of the presence or absence of the mass-inflation instability for regular black hole geometries at the classical level is reasonably complete. The main conclusion is that the mass-inflation instability established for the Reisner-Nordstr{\"o}m geometry does not automatically extend to regular black holes. This raises the intriguing question whether the failure of this analogy extends to the quantum level. This may be investigated along the lines \cite{Casals:2016odj, Casals:2018ohr, Sela:2018xko, Casals:2019jfo, Zilberman:2019buh, Klein:2021ctt,  Zilberman:2022aum, Klein:2024sdd, Arrechea:2024ajt}, using techniques developed in the context of quantum field theory in a curved spacetime. We hope to come back to this question in the future.

\section*{Acknowledgements}
A.\ P.\ thanks Francesco Di Filippo and Jesse van der Duin for useful discussions. A.\ P.\ acknowledges partial financial support by Erasmus+ grants no.\ 2022-1-IT02-KA131-HED-000066529 and no.\ 2023-1-IT02-KA131-HED-000135582.
\appendix
\section{Curvature Invariants}
\label{App.A}
The geometries studied in this work are of the type
\be
ds^2=-f(v,r)dv^2+2drdv+r^2d\Omega^2 \, , 
\ee
where
\be\label{lapse1}
f(v,r)=1-\frac{2M(v,r)}{r} \, .
\ee
In order to quantify the curvature of spacetime, we introduce the Riemann tensor $R_{\alpha \beta \gamma \delta}$, the Ricci tensor $R_{\mu\nu} = R^\alpha{}_{\mu\alpha\nu}$ and the Ricci scalar $R \equiv g^{\mu \nu}R_{\mu \nu}$. Furthermore, we use a prime to denote derivatives with respect to $r$, i.e., $M'(r,v)\equiv \partial M(r,v)/\partial r$.

For the geometry \eqref{lapse1} the Ricci scalar can be given in terms of the Misner-Sharp mass as
\be\label{Ricci-scalar}
R=\frac{2rM''+4M'}{r^2} \, ,
\ee
In addition, we introduce the two curvature scalars at quadratic order in the spacetime curvature. The Kretschmann scalar $K\equiv R_{\alpha \beta \gamma \delta}R^{\alpha \beta \gamma \delta}$ evaluates to 
\begin{align}\label{def:K}
K = &\frac{48M^{2}}{r^6}-\frac{64MM'}{r^5}+\frac{32(M')^{2}}{r^4} \\ \notag
  &+\frac{16MM''}{r^4}-\frac{16M'M''}{r^3}+\frac{4(M'')^2}{r^2} \, , 
\end{align}
 and the square of the Weyl-tensor $C^{2} \equiv C_{\alpha \beta \gamma \delta}C^{\alpha \beta \gamma \delta}$ reads
\begin{align} \label{Weyl}
  C^2=&\frac{4(r^2M''-4rM'+6M)^2}{3r^6} \, .
\end{align}
This two quantities are related by the identity
\be
K=C^2-2R_{\mu \nu}R^{\mu \nu}+\frac{1}{3} R^2 \, . 
\ee
Since the Reissner-Nordstr{\"o}m solution and the regular black hole solutions are not Ricci flat ($R_{\mu \nu} \ne 0$), they describe two different aspects of spacetime and in principle could exhibit different behaviors as consequence of the perturbation. Finally, we also introduce the Coulomb-component $\Psi_{2}$. In the coordinate systems used in this work $\Psi_{2} = -\frac{1}{2}C^{\theta \varphi}{}_{\theta \varphi}$. Written in terms of the Misner-Sharp mass one then has 
\be\label{def:psi2}
\Psi_{2}=-\frac{M}{r^3}+\frac{2M'}{3r^2}-\frac{M''}{6r} \, .
\ee
Both $C^2$ and $\Psi_2$ give measures for the tidal forces experienced by a massive object.
\section{Attractor for the AS-collapse}
\label{App.D}
This appendix provides the details underlying the derivation that the late-time attractor \eqref{mp.rep2.AScollapse} induces the logarithmic divergence of the Misner-Sharp mass given in \eqref{result.main}.

We start from eq.\ \eqref{eq.mp.collapse} which encodes the dynamics of $m_+(v)$. The late-time attractor \eqref{mp.att1.AScollapse} is characterized by
\be
\left( 1 + \frac{6 \xi m_+}{R^3} \right) \ll 1 \, .
\ee
This motivates introducing the new function
\be
x(v) \equiv 1 + \frac{6 \xi m_+(v)}{R(v)^3} \, . 
\ee
Recasting eq.\ \eqref{eq.mp.collapse} in terms of this new function yields
\be\label{mp-to-x}
\dot{x} = 3 \left( 1 - x \right) \frac{\dot{R}}{R} + \frac{6 \xi}{R^3} \, x \, \left(1 - \frac{R^2}{6 \xi} \log x^2 \right) \, F(v) \, . 
\ee
At this stage, no approximation has been made. Subsequently, we approximate $R$, $\dot{R}$, and $F(v)$ by their leading late-time behavior
\be
R \simeq r_- \, , \quad \dot{R} \simeq (1-p) \frac{a_0}{v^p} \, , \quad F(v) \simeq - \frac{r_- \kappa_-}{2} \, , 
\ee
with $a_0$ given in eq.\ \eqref{a0coeff}. Introducing the abbreviations
\be\label{ab.def}
\begin{split}
a = - \frac{(p-1) \beta}{\kappa_- r_-^3} \left( 1 + \frac{6 m_0 \xi}{r_-^3} \right)^{-1} \, , \; \;  b \equiv - \frac{3 \xi \kappa_-}{r_-^2} \, , 
\end{split}
\ee
the equation encoding the late-time dynamics of $x(v)$ reads
\be\label{x-latetime}
\dot{x} = 3 a v^{-p} + b \, x + \frac{\kappa_-}{2} \, x \, \log(x   ^2) \, . 
\ee
Here the term proportional to $x \dot{R}$ has been dropped since it does not contribute to the leading behavior at late times. Unfortunately, we are not aware of an analytic solution of \eqref{x-latetime}.

In order to progress further, we decompose
\be 
x(v) = \frac{1}{v^t} \, y(v) \, ,
\ee 
where $t$ is a free coefficient capturing the power-law scaling of $x(v)$. We then recast \eqref{x-latetime} in terms of the scaling operator associated with $y(v)$:
\be\label{y-latetime}
\begin{split}
\frac{d \log y}{d \log v} \, = & \, t  + 3 a \, y^{-1} \, v^{t+1-p} + b v \\ &  + \frac{\kappa_-}{2} \, v \log y^2 - t \, \kappa_- \, v \log v \, .
\end{split} 
\ee

At this point, let us assume that $y(v) = y_0$ is a $v$-independent constant. This is equivalent to the assumption that $x(v)$ exhibits a power-law behavior at late times. Under this assumption, the left-hand side of \eqref{y-latetime} vanishes. Requiring that the terms containing just powers of $v$ on the right-hand side cancel then fixes
\be\label{t-fix}
t = p \, , 
\ee
and 
\be\label{y0.sol}
a + b\ y_0 + \frac{1}{2} \kappa_- \, y_0 \log(y_0)^2 = 0 \, .
\ee

The assumption $y(v) = y_0$ cannot be entirely correct, though, since it does not account for the $v \log v$-term appearing in the last line of eq.\ \eqref{y-latetime}. These contributions can be consistently accounted for by generalizing our initial assumption to
\be\label{y-ansatz}
y(v) = y_0 \, \left[ 1 + \sum_{n=1} \, y_n \, \log^n(v) \right] \, , 
\ee
where $y_n$ are free coefficients. The new terms in \eqref{y-ansatz} then induce logarithmic corrections to the power-law scaling of $x(v)$. Substituting \eqref{y-ansatz} into \eqref{y-latetime} we arrive at
\be\label{yt-latetime}
\begin{split}
\frac{d \log \tilde{y}}{d \log v} \, = & \, 12 + \frac{3 a}{y_0} \, v \, \left( \frac{1}{\tilde y} - 1 \right) \\ & \, + \kappa_- \, v \, \log \tilde{y} - 12 \kappa_- v \log v \, .
\end{split} 
\ee
Here, we abbreviated $\tilde{y} \equiv  \left[ 1 + \sum_{n=1} \, y_n \, \log^n(v) \right]$ and used \eqref{y0.sol} to eliminate the terms linear in $v$. Notably, the left-hand side and the first term on the right-hand side of \eqref{yt-latetime} are subleading in the late-time dynamics and can be dropped. The remainder of the equation reads
\be\label{gen-functional}
\frac{3 a}{y_0} \, v \, \left( \frac{1}{\tilde y} - 1 \right) + \kappa_- \, v \, \log \tilde{y} - 12 \kappa_- v \log v = 0 \, .
\ee
This equation has the status of a generating functional that fixes the free coefficients $y_n$. Expanding in powers of $\log v$, one obtains a hierarchy of equation that allows to fix all $y_n$ recursively, by solving linear equations. The $v \log v$-term thereby serves as a seed that excludes the trivial solution. The structure of \eqref{gen-functional} also shows that the generalization \eqref{y-ansatz} is sufficient to capture the late-time dynamics of the attractor.

When reconstructing the Misner-Sharp mass $M_+(v)$, one then takes into account that inserting the product \eqref{y-ansatz} into the logarithm turns the bracket into an additive (and subleading) contribution. In this way, one establishes that \eqref{result.main} provides a well-founded approximation of the late-time dynamics associated with the attractive quasi-fixed point of the model.

\twocolumngrid
\nocite{*}

\end{document}


\begin{tikzpicture}
    \coordinate (H) at (0,0);
    \coordinate (C) at (-4,4);
    \coordinate (E) at (-4,-4);
    \coordinate (I) at (4,-4);
    \coordinate (F) at (0,4);

    \draw[thick] (H)--(E);
    \draw[thick] (I) -- (H);
   \draw[very thick, dashed] (C) -- (-1.15,1.15); 
    \draw[thick] (-1.15,1.15) -- (H); 

    \draw[thick, decorate, decoration={zigzag, segment length=2mm, amplitude=0.5mm}] (H) -- (F);

    \draw[thick, ->, decorate, decoration={snake, segment length=1mm, amplitude=0.5mm}] (0,-1) -- (-2-1,1+1);
    \draw[thick, ->, decorate, decoration={snake, segment length=2mm, amplitude=0.5mm}] (-0.5,-1.5) -- (-2.5-1,0.5+1);
    \draw[thick, ->, decorate, decoration={snake, segment length=3mm, amplitude=0.5mm}] (-1,-2) -- (-3-1,0+1);

    \draw[very thick, blue] (-4,-1.7) -- (-0.5,1.8); 

    \draw[thick] (-0.5,1.8) -- (-2.5,3.8); 

    \node[above, rotate=-45] at (-1.84,3.16) {$f_+=0$};

 \node[above, rotate=-45] at (-3.15,3.15) {$v_+=V=0$};

    \node at (C) [left] {$\text{C}$};
    \node at (E) [left] {$\text{E}$};
    \node at (I) [right] {$\mathcal{I}^+$};
    \node at (H) [above right] {$\text{H}$};
    \node[rotate=-45] at (-0.6,0.9) {$f_{-}=0$};  
    \node at (0.5,2) {$r = 0$};
    \node[rotate=-45] at (1.5,-1.2) {$v_- = v = \infty$};
    \node at (-3.5,-.5) [circle, draw, inner sep=1pt] {$+$};
    \node at (-2.9,-1.1) [circle, draw, inner sep=1pt] {$-$};
    \node at (-3.8,-1.1) {$\Sigma$};
\end{tikzpicture}